\documentclass{iopart}
\usepackage{iopams}

\usepackage{graphicx}
\usepackage{subfigure}
\usepackage{lscape}
\usepackage[vcentermath]{youngtab}

\newcommand{\vev}[1]{\langle #1\rangle}
\newcommand{\bvev}[1]{\bigl\langle #1 \bigr\rangle}
\newcommand{\bra}[1]{\langle #1 |}
\newcommand{\ket}[1]{| #1 \rangle}
\newcommand{\braket}[2]{\langle #1 | #2\rangle}

\newcommand{\bket}[1]{\bigl| #1 \bigr\rangle}

\newcommand{\0}{\underline{0}}
\newcommand{\1}{\underline{1}}
\newcommand{\2}{\underline{2}}
\newcommand{\3}{\underline{3}}
\newcommand{\4}{\underline{4}}
\newcommand{\uom}{\underline{\omega}}

\begin{document}

\title[Spin and Rotations in Galois Field Quantum Mechanics]{Spin and Rotations in\\Galois Field Quantum Mechanics}

\author{Lay Nam Chang, Zachary Lewis,\\Djordje Minic and Tatsu Takeuchi}

\address{Department of Physics, Virginia Tech, Blacksburg, VA 24061, USA}

\eads{\mailto{laynam@vt.edu},\mailto{zlewis@vt.edu},\mailto{dminic@vt.edu},\mailto{takeuchi@vt.edu}}

\begin{abstract}
We discuss the properties of Galois Field Quantum Mechanics constructed
on a vector space over the finite Galois field $GF(q)$.
In particular, we look at 2-level systems analogous to spin, and
discuss how $SO(3)$ rotations could be embodied in such a system.
We also consider two-particle `spin' correlations and show that the
Clauser-Horne-Shimony-Holt (CHSH) inequality is nonetheless not violated in this model.
\end{abstract}

\pacs{03.65.Aa, 03.65.Ta, 03.65.Ud}
\submitto{\JPA}

\maketitle
\section{Introduction}

Correlations of physical properties between states provide the cleanest differentiation between quantum descriptions and classical ones.   
By quantum descriptions we mean those in which entangled states have correlations that cannot be reproduced by ascribing to the physical properties
distinctive and unchanging values.   
By classical descriptions we mean those wherein such assignments are explicit, or through hidden variables.   The best known of this distinction is encoded in the super-classical descriptions of spins in canonical quantum mechanics (QM), 
as exemplified by the celebrated Bell's inequalities \cite{bell}.

In the formulation by Clauser, Horne, Shimony, and Holt (CHSH) \cite{Clauser:1969ny},
the observables $A_\alpha$ and $B_\beta$,
associated with the spins of two spacelike separated particles and 
the subscripts indicating detector settings, 
are assumed to yield either one of the two numbers $\pm 1$ upon measurement.
The combination of correlators given by
\begin{equation}
\bvev{A_1,A_2;B_1,B_2}\,\equiv\,
\bvev{A_1 B_1}+\bvev{A_1 B_2}+\bvev{A_2 B_1}-\bvev{A_2 B_2}\;,
\label{CHSHcorrdef}
\end{equation}
can be shown to have its absolute value bounded by 2 for
classical hidden variable theory, that is:
\begin{equation}
\Bigl|\bvev{A_1,A_2;B_1,B_2}\Bigr|\;\le\;2\;,
\label{CHSHbound}
\end{equation}
for arbitrary pairs of detector settings.
This inequality is violated by correlations in canonical QM
in which the upper bound is replaced by $2\sqrt{2}$ \cite{cirelson,Chang:2011yt}. 
In the present context, by canonical QM we mean a description based upon a Hilbert space defined over the complex number field.  Observables have correspondence with hermitian operators on the space, which do not necessarily commute with each other.  

However, we demonstrate in \cite{Chang:2012} that canonical QM is not
a unique quantum description.
There, variants of canonical QM are constructed on vector spaces over the finite Galois field $GF(q)$. 
Both canonical QM and our variant quantum description make crucial use of the underlying linear properties, and the CHSH bound for both is predicated upon properties of the vector space.
In \cite{Chang:2011yt} we point out the importance of the inner product in deriving the conventional CHSH bound.   
The vector spaces on which our variants are constructed
do not support an inner product because of the cyclicity of Galois fields.  
Consequently, the CHSH bound of our variant differs from the canonical value:
we find that it is 2, the classical hidden variable value.  
Nevertheless, we show that  the resultant range of correlations also cannot be reproduced in any classical context with hidden variables \cite{Chang:2012}.   The system is therefore quantum in the sense described above, despite the CHSH bound remaining at 2.

The use of Galois fields in physics is not extensive \cite{nambu}.
In this paper, we elaborate on our model presented in \cite{Chang:2012}
to better provide an intuitive feel for how the properties of
Galois fields are utilized. 
Details of implementation using concrete examples are presented,
as well as the full derivation of the CHSH bound for `spin' systems.

Our model system for `spin' has much similarity with canonical spin in conventional 3D space.   
We describe the extent to which our `spin' can be mapped to ordinary spin, 
and the associated finite projective geometry and projective group to 3D space rotations.     
This relationship is demonstrated via transformations on the associated polyhedron. 
The emergent geometry provides a useful framework to visualize the underlying physics of the approach.  
In general, the role of $SU(N)$ familiar in descriptions of spin-like $N$-level systems in conventional QM is now replaced by $PGL(N,q)$, the projective linear group acting on the vector space over Galois fields of order $q$.

This paper is organized as follows.  
In section II, we review our construction of Galois field quantum mechanics (GQM).  
Section III presents a detailed analysis of the $GF(2)$ case. 
We pay particular attention to the CHSH bound, 
and the impossibility of hidden variables to account for this result.  We also present the analogs of singlet and triplet states under $SU(2)$ in conventional QM in the present formulation, and how these account for the resultant bound.  
Section IV is devoted to the discussion of the geometric structure of `rotations' in 
the contexts of $GF(3)$, $GF(4)$ and $GF(5)$ fields.   
We explicitly construct and identify some of the transformations with those of some polyhedral group, for the case of $N=2$ and $q=2$, $3$, $4$, and $5$,  thereby justifying the spin-like context for these levels.    
In the concluding section, we comment on how these results can impact upon more 
general foundational questions in quantum theory.

Before we proceed, we emphasize that our model is distinct from `Galois quantum systems' discussed in the literature \cite{Vourdas:04, Vourdas:07}.  These papers consider a phase space which is assumed to be $GF(q) \times GF(q)$; that is, the position and momentum of particles take values in $GF(q)$.   In this paper, it is the wave-functions that take values in $[GF(q)]^N$, while the outcomes of measurements take on values in $\mathbb{R}$, the real number line.

\section{Galois Field Quantum Mechanics}

The key variation introduced in \cite{Chang:2012} is the replacement of
the Hilbert space of an $N$-level quantum system, $\mathcal{H}_\mathbb{C}=\mathbb{C}^N$,
with a discrete vector space over a finite field: $\mathcal{H}_q=\mathbb{Z}_q^N$ \cite{MQT,Finkelstein:1996fn}.
Here, $\mathbb{Z}_q$ denotes the Galois field $GF(q)$, where
$q=p^n$ with $p$ prime and $n\in\mathbb{N}$.
For the $n=1$ case, $GF(p)$ is simply $\mathbb{Z}_p=\mathbb{Z}/p\mathbb{Z}$.
The states of the system are represented by vectors $\ket{\psi}\in\mathcal{H}_q$, 
and outcomes of measurements by dual-vectors $\bra{x}\in\mathcal{H}_q^*$.
Observables are associated with a choice of basis of $\mathcal{H}_q^*$,
each dual-vector in it representing a different outcome.
The probability of obtaining the outcome represented by $\bra{x}$
when a measurement of the observable is performed on the state represented by $\ket{\psi}$
is given by the canonical form
\begin{equation}
P(x|\psi) \;=\; \frac{\bigl|\braket{x}{\psi}\bigr|^2}{\sum_y \bigl|\braket{y}{\psi}\bigr|^2}\;,
\label{Pdef}
\end{equation}
where the bracket $\braket{x}{\psi}\in\mathbb{Z}_q$ is converted into
a non-negative real number $|\braket{x}{\psi}|\in\mathbb{R}$ via the absolute value function:
\begin{equation}
|\,\underline{k}\,|\;=\;
\left\{\begin{array}{ll}
0\quad &\mbox{if $\underline{k}=\0$}\;,\\
1\quad &\mbox{if $\underline{k}\neq\0$}\;.
\end{array}
\right.
\label{abs}
\end{equation}
Here, underlined numbers and symbols represent elements of $\mathbb{Z}_q$, to
distinguish them from elements of $\mathbb{R}$ or $\mathbb{C}$.
Note that all non-zero elements of $\mathbb{Z}_q$ are assigned an absolute value
of 1, effectively making them all `phases.'

Definition \eref{abs} is not arbitrary.   In order that the absolute value function 
generate probability amplitudes for multi-particle states, we must ensure, for consistency, that it satisfies the critical factorizability criterion for product states.    
Since $\mathbb{Z}_q\backslash\{\0\}$ is a cyclic multiplicative group,
this definition of absolute value is the only one consistent with the requirement
\begin{equation}
|\underline{k}\underline{l}|\;=\;|\underline{k}||\underline{l}|\;,
\label{ProductPreservingMap}
\end{equation}
which is necessary for the probabilities of product states to factorize.   

Since the multiplication of $\ket{\psi}$ with a non-zero element of
$\mathbb{Z}_q$ will not affect the probability as defined above, 
vectors that differ by a non-zero multiplicative constant
are identified as representing the same 
physical state, and the state space is endowed with the finite projective geometry
\cite{Hirschfeld,Arnold,Ball-Weiner}
\begin{equation}
PG(N-1,q) 
\;=\; (\,\mathbb{Z}_q^N\backslash\{\mathbf{\0}\}\,)\,\big/\,(\,\mathbb{Z}_q\backslash\{\0\}\,) \;.
\end{equation}
The group of all possible basis transformations in this space is 
the projective group $PGL(N,q)$:
\begin{equation}
PGL(N,q)\;=\;GL(N,q)\,\big/\,Z(N,q)\;.
\end{equation}
Here, $GL(N,q)$ is the general linear group of $\mathbb{Z}_q^N$, and
$Z(N,q)$ is its center, which consists of $N\times N$ unit matrices
multiplied by a `phase' in $\mathbb{Z}_q\backslash\{\0\}$.
Effectively, the elements of $PGL(N,q)$ lead to permutations of
the states in $PG(N-1,q)$, and is thus a subgroup of the 
symmetric group of all possible state permutations.
Physically, basis transformations should correspond to the change of
`detector setting,' \textit{e.g.} the rotation of the polarizaton
axis of a spin-measuring device.

Let us denote the GQM model resulting from this procedure as $GQM(N,q)$.
Spin-like systems with two possible outcomes $\pm 1$ can be constructed
on the space $V_q\equiv\mathbb{Z}_q^2$ as $GQM(2,q)$, and two-particle spin-like systems
on $V_q\otimes V_q=\mathbb{Z}_q^2\otimes\mathbb{Z}_q^2=\mathbb{Z}_q^4$ as $GQM(4,q)$.
In the following, we will consider the cases $q=2$, $3$, $4$, and $5$ 
as concrete examples of this procedure.

\section{$\mathbb{Z}_2$ Quantum Mechanics}

\subsection{One-Particle Spin}

We begin our discussion with $q=2$.
The spin-like system is $GQM(2,2)$,
perhaps the simplest quantum system imaginable, which is constructed on
the finite field consisting of only two elements, $GF(2)=\mathbb{Z}_2=\mathbb{Z}/2\mathbb{Z}=\{\0,\1\}$,
with the following addition and multiplication tables:
\begin{center}
\begin{tabular}[t]{c|cc}
$\;+\;$ & $\;\0\;$ & $\;\1\;$ \\
\hline
$\0$ & $\0$ & $\1$ \\
$\1$ & $\1$ & $\0$ 
\end{tabular}
\qquad
\begin{tabular}[t]{c|cc}
$\;\times\;$ & $\;\0\;$ & $\;\1\;$ \\
\hline
$\0$ & $\0$ & $\0$ \\
$\1$ & $\0$ & $\1$ 
\end{tabular}
\end{center}
As stated above, we use numbers with underlines to indicate elements of $\mathbb{Z}_2$
to distinguish them from elements of $\mathbb{R}$ or $\mathbb{C}$.
Since $\mathbb{Z}_2\backslash\{\0\}=\{\1\}$ is trivial, there are no phases 
and each physical state is represented by a unique non-zero vector in $V_2=\mathbb{Z}_2^2$.

There exist only $2^2-1=3$ non-zero vectors in $V_2$, 
which we denote:
\begin{equation}
\ket{a}\,=\,\left[\begin{array}{c} \1 \\ \0 \end{array}\right]\,,\quad
\ket{b}\,=\,\left[\begin{array}{c} \0 \\ \1 \end{array}\right]\,,\quad
\ket{c}\,=\,\left[\begin{array}{c} \1 \\ \1 \end{array}\right]\,.
\label{VectorRep}
\end{equation}
Thus, there are 3 possible states of the system.
Since
\begin{eqnarray}
\ket{a} \;=\; \ket{b}+\ket{c}\;,\quad
\ket{b} \;=\; \ket{c}+\ket{a}\;,\quad
\ket{c} \;=\; \ket{a}+\ket{b}\;,
\end{eqnarray}
%
the three vectors are completely equivalent;
any pair of them can be used as a basis for $V_2$.
A change of basis in $V_2$ would permute the above
column representations among the three vectors,
or equivalently, permute the vector
labels $a$, $b$, and $c$ on the three column vectors.
Since all permutations of the three vectors are possible, 
the group of basis transformations on $V_2$ is $S_3\cong PGL(2,2)$.

The dual vector space $V_2^*$ is the space of all linear maps
from $V_2$ to $\mathbb{Z}_2$.
There are $2^2-1=3$ non-zero dual vectors in $V_2^*$, 
and following \cite{Chang:2012}, we denote them as:
\begin{equation}
\bra{\bar{a}} \;=\; \bigl[\;\0\;\;\1\;\bigr]\;,\quad
\bra{\bar{b}} \;=\; \bigl[\;\1\;\;\0\;\bigr]\;,\quad
\bra{\bar{c}} \;=\; \bigl[\;\1\;\;\1\;\bigr]\;.
\end{equation}
%
This labeling allows us to write:
\begin{equation}
\braket{\bar{r}}{s} \;=\;
\left\{ \begin{array}{ll}
\0 \quad & \mbox{if $r=s$,}\\
\1       & \mbox{if $r\neq s$,}
\end{array}
\right.
\end{equation}
and consequently,
\begin{equation}
|\braket{\bar{r}}{s}| \;=\; 1-\delta_{rs}\;.
\label{antiorthogonality}
\end{equation}
This relation pairs up vectors with dual-vectors in a somewhat non-standard way,
and leads to the various consequence of GQM.
To maintain it, we assume that a relabeling of vectors in $V_2$ is always accompanied
by a corresponding relabeling of dual-vectors in $V_2^*$.

Observables are associated with \textit{a choice of basis} in $V_2^*$.
There are six possible choices:
\begin{equation}
A_{rs}\;=\;\{\,\bra{\bar{r}},\,\bra{\bar{s}}\,\}\;,
\label{Apqdef}
\end{equation}
with $rs=ab$, $ba$, $bc$, $cb$, $ca$, and $ac$.
Each of the dual vectors in each basis represents 
an \textit{outcome}\footnote{%
Schumacher and Westmoreland call it an \textit{effect} in \cite{MQT}} 
which could occur as the result of a measurement of the observable
represented by that basis,
that is, the measurement of the observable $A_{rs}$
would result in one of the two outcomes represented by $\bra{\bar{r}}$ and $\bra{\bar{s}}$.

Though elements of $V_2^*$ map elements of $V_2$ onto $\mathbb{Z}_2$, the outcomes they represent 
need not be elements of $\mathbb{Z}_2$ themselves.
We assign the numerical values $\pm 1$ to the two outcomes 
represented by the pair of dual vectors in each basis:
$+1$ to the first dual vector, and $-1$ to the second dual vector.
Thus, $\bra{\bar{r}}$ represents the outcome $+1$ when $A_{rs}$ is measured,
while $\bra{\bar{s}}$ represents the outcome $-1$ when $A_{rs}$ is measured.
If the ordering of the dual vectors is reversed to 
$\{\,\bra{\bar{s}},\,\bra{\bar{r}}\,\}$, then this pair corresponds to $-A_{rs}$. 
Thus $A_{sr}=-A_{rs}$, and we can consider $A_{rs}$ and $A_{sr}$ to be essentially the
same observable.  So the number of observables in our system is three:
$A_{ab}$, $A_{bc}$, and $A_{ca}$.

\begin{figure}[b]
\begin{indented}
\item[]
\includegraphics[height=5cm]{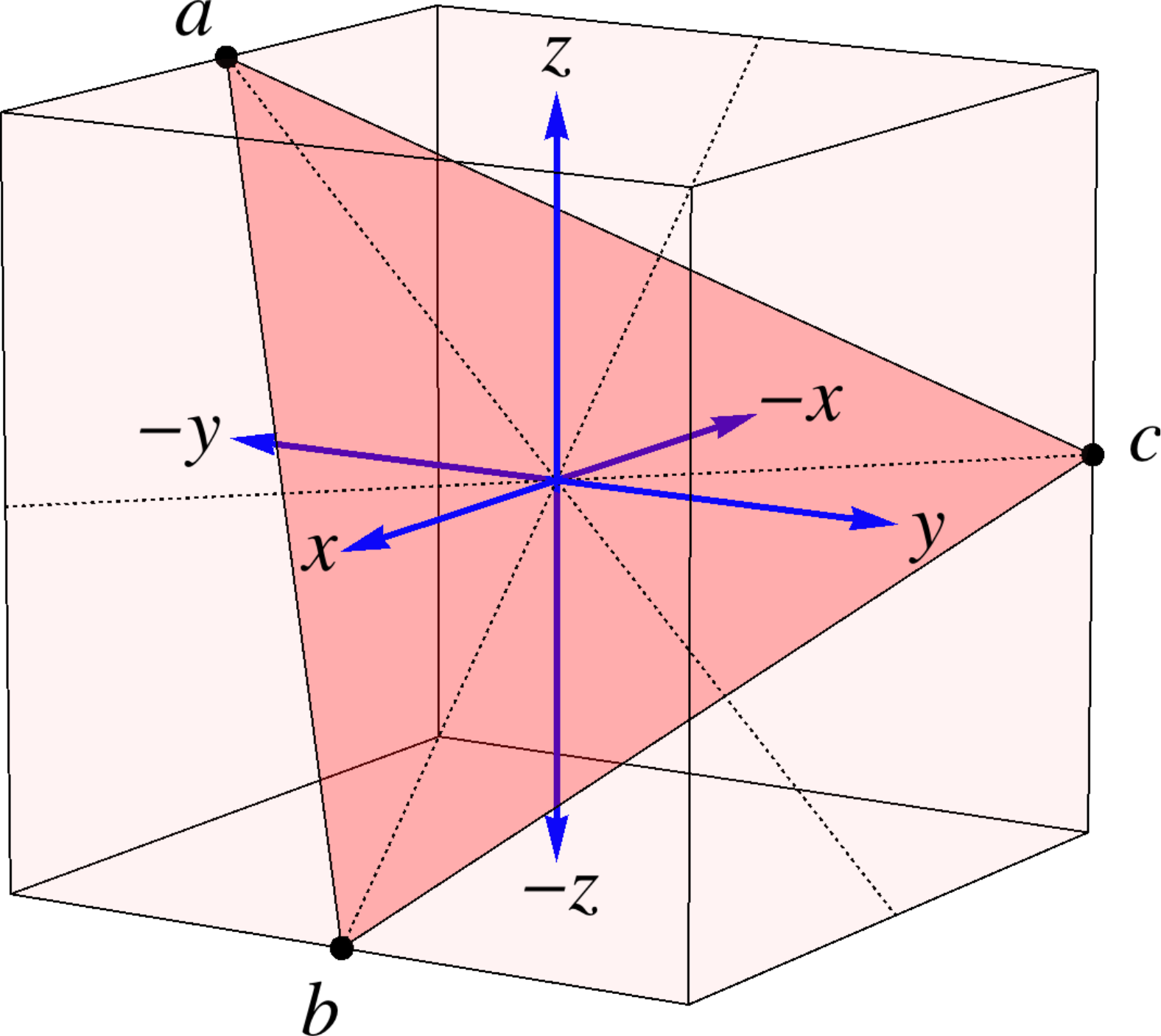}
\end{indented}
\caption{\label{Fig1}The 6 `spin' directions of $GQM(2,2)$.
Allowed $SO(3)$ rotations are those that rotate the equilateral triangle $abc$ onto itself.\\}
\end{figure}

The above assignment of outcomes allows us to view $A_{rs}$ as representing 
observables akin to `spin,' with the indices $rs$ representing the
`direction' of the spin.   
\cite{MQT} labels them as:
\begin{equation}
X\;=\;A_{bc}\;,\qquad Y\;=\;A_{ca}\;,\qquad Z\;=\;A_{ab}\;.
\end{equation}
Indeed, if we look at their transformation properties under the $S_3$ group
of basis transformations, we find:
\begin{equation}
\hspace{-0.5cm}
\begin{array}{rlrlrl}
(ab)X &\! = \, -Y\,,\quad & 
(ab)Y &\! = \, -X\,,\quad &
(ab)Z &\! = \, -Z\,,\\
(bc)X &\! = \, -X\,,\quad &
(bc)Y &\! = \, -Z\,,\quad &
(bc)Z &\! = \, -Y\,,\\
(ca)X &\! = \, -Z\,,\quad &
(ca)Y &\! = \, -Y\,,\quad &
(ca)Z &\!= \, -X\,,\\
(abc)X &\!= \, +Y\,, & 
(abc)Y &\!= \, +Z\,, &
(abc)Z &\!= \, +X\,,\\
(acb)X &\!= \, +Z\,, & 
(acb)Y &\!= \, +X\,, &
(acb)Z &\!= \, +Y\,,
\end{array}
\label{braPermutations}
\end{equation}
which can all be considered $SO(3)$ rotations of the mutually orthogonal
$X$, $Y$, and $Z$ axes as shown in \fref{Fig1}.
The six basis transformations in $S_3$ can be mapped onto the
six rotations of the dihedral group $D_3$ which keep the equilateral triangle $abc$
in \fref{Fig1} invariant.

Thus, our `spins' transform in an analogous way to canonical spin
under `rotations.'
However, a significant difference also exists.
With our `spin,' the same dual vector can represent different outcomes of different 
observables: in addition to the outcomes of $A_{ab}$, 
$\bra{\bar{a}}$ represents the outcome $-1$ when $A_{ca}$ is measured,
while $\bra{\bar{}b}$ represents the outcome $+1$ when $A_{bc}$ is measured.
What each dual vector represents depends on the observable under consideration.

The probabilities of outcomes are calculated with \eref{Pdef}.
For the measurement of the observable $Z=A_{ab}$, for instance, 
we find:
\begin{eqnarray}
P(A_{ab}\,;+\,|\,a) 
& = & 
\frac{\bigl|\braket{\bar{a}}{a}\bigr|^2}
{\bigl|\braket{\bar{a}}{a}\bigr|^2+\bigl|\braket{\bar{b}}{a}\bigr|^2}
\;=\;
0
\;,
\cr
P(A_{ab};-\,|\,a) 
& = & 
\frac{\bigl|\braket{\bar{b}}{a}\bigr|^2}
{\bigl|\braket{\bar{a}}{a}\bigr|^2+\bigl|\braket{\bar{b}}{a}\bigr|^2}
\;=\;
1
\;,
\cr
P(A_{ab};+\,|\,b) 
& = & 
\frac{\bigl|\braket{\bar{a}}{b}\bigr|^2}
{\bigl|\braket{\bar{a}}{b}\bigr|^2+\bigl|\braket{\bar{b}}{b}\bigr|^2}
\;=\;
1
\;,
\cr
P(A_{ab};-\,|\,b) 
& = & 
\frac{\bigl|\braket{\bar{b}}{b}\bigr|^2}
{\bigl|\braket{\bar{a}}{b}\bigr|^2+\bigl|\braket{\bar{b}}{b}\bigr|^2}
\;=\;
0
\;,
\cr
P(A_{ab};+\,|\,c) 
& = & 
\frac{\bigl|\braket{\bar{a}}{c}\bigr|^2}
{\bigl|\braket{\bar{a}}{c}\bigr|^2+\bigl|\braket{\bar{b}}{c}\bigr|^2}
\;=\;
\frac{1}{2}
\;,
\cr
P(A_{ab};-\,|\,c) 
& = & 
\frac{\bigl|\braket{\bar{b}}{c}\bigr|^2}
{\bigl|\braket{\bar{a}}{c}\bigr|^2+\bigl|\braket{\bar{b}}{c}\bigr|^2}
\;=\; 
\frac{1}{2}
\;.\quad
\end{eqnarray}
The expectation values of $Z=A_{ab}$ on the three states are therefore:
\begin{equation}
\begin{array}{llll}
\vev{A_{ab}}_a & =\; (+1)\times 0 & +\;(-1)\times 1 & =\; -1\;,\\
\vev{A_{ab}}_b & =\; (+1)\times 1 & +\;(-1)\times 0 & =\; +1\;,\\
\vev{A_{ab}}_c & =\; (+1)\times\frac{1}{2} & +\;(-1)\times\frac{1}{2} & =\; \phantom{+}0\;.
\end{array}
\label{Xvev}
\end{equation}
Thus, $\ket{a}$ and $\ket{b}$ take on the role of the `eigenstates' of $A_{ab}$, while
$\ket{c}$ is the superposition of the two with a 50-50 chance of obtaining either
$+1$ or $-1$.
The probabilities and expectation values of all other observables can be calculated in a similar fashion
and we obtain the results listed in \tref{oneparticlevevs}.

The probabilities for $X=A_{bc}$ and $Y=A_{ca}$ can also be calculated by `rotating' the results for
$Z=A_{ab}$ via \eref{braPermutations}.
For instance, since $(abc)A_{ab}=A_{bc}$, we can conclude that
\begin{eqnarray}
P(A_{bc};+\,|\,b) & = & (abc)\,P(A_{ab};+\,|\,a) \;=\; 0\;, \cr
P(A_{bc};-\,|\,b) & = & (abc)\,P(A_{ab};-\,|\,a) \;=\; 1\;.
\end{eqnarray}
%

\begin{table}[t]
\caption{\label{oneparticlevevs}
The probabilities of the two outcomes $+$ and $-$ for all combinations
of observables and states in $GQM(2,2)$.}
\begin{indented}\item[]
\begin{tabular}{ccccc}
\br
\ observable\ \ &\ state\ \ &\ $P(+)$\ \ &\ $P(-)$\ \ & Expectation Value \\
\br
    & $a$ & $0$           & $1$           & $-1$ \phantom{\big|} \\
\cline{2-5}
$A_{ab}$ 
	& $b$ & $1$           & $0$           & $+1$ \phantom{\big|} \\
\cline{2-5}
    & $c$ & $\frac{1}{2}$ & $\frac{1}{2}$ & $\phantom{-}0$ \phantom{\big|} \\
\br
    & $a$ & $\frac{1}{2}$ & $\frac{1}{2}$ & $\phantom{-}0$ \phantom{\big|}\\
\cline{2-5}
$A_{bc}$
    & $b$ & $0$           & $1$           & $-1$ \phantom{\big|}\\
\cline{2-5}
    & $c$ & $1$           & $0$           & $+1$ \phantom{\big|}\\
\br
    & $a$ & $1$           & $0$           & $+1$ \phantom{\big|}\\
\cline{2-5}
$A_{ca}$ 
	& $b$ & $\frac{1}{2}$ & $\frac{1}{2}$ & $\phantom{-}0$ \phantom{\big|}\\
\cline{2-5}
    & $c$ & $0$           & $1$           & $-1$ \phantom{\big|}\\
\br
\end{tabular}
\end{indented}
\end{table}

Note that, due to our construction, states $\ket{r}$ and $\ket{s}$ are `eigenstates' of $A_{rs}$
for all $rs$.
However, in our approach, observables are not linear hermitian maps from $V_2$ to $V_2$
as in canonical QM, and thus do not have eigenstates in the usual
sense of the term.

Note, also, that each state is an `eigenstate' of two observables at a time:
$\ket{a}$ of $A_{ab}$ and $A_{ca}$, $\ket{b}$ of $A_{bc}$ and $A_{ab}$, and $\ket{c}$ of $A_{ca}$ and $A_{bc}$.
This is due to each of the three dual vectors appearing in two observables.
Thus, despite the resemblance to spins in the $x$, $y$, and $z$ directions
in canonical QM, this quantum system is quite distinct.

\subsection{Two Particle States}

Two particle states are expressed as vectors in the
tensor product space $V_2\otimes V_2 = \mathbb{Z}_2^4$, and the system becomes $GQM(4,2)$.
There are $2^4-1=15$ non-zero vectors in this space, of which nine
are product states and six are entangled states.
The nine product states are:
\begin{equation}
\begin{array}{lll}
\ket{aa} & = \; \ket{a}\otimes\ket{a} & =\; \bigl[\;\1\;\;\0\;\;\0\;\;\0\;\bigr]^\mathrm{T}\;, \\
\ket{ab} & = \; \ket{a}\otimes\ket{b} & =\; \bigl[\;\0\;\;\1\;\;\0\;\;\0\;\bigr]^\mathrm{T}\;, \\
\ket{ac} & = \; \ket{a}\otimes\ket{c} & =\; \bigl[\;\1\;\;\1\;\;\0\;\;\0\;\bigr]^\mathrm{T}\;, \\
\ket{ba} & = \; \ket{b}\otimes\ket{a} & =\; \bigl[\;\0\;\;\0\;\;\1\;\;\0\;\bigr]^\mathrm{T}\;, \\
\ket{bb} & = \; \ket{b}\otimes\ket{b} & =\; \bigl[\;\0\;\;\0\;\;\0\;\;\1\;\bigr]^\mathrm{T}\;, \\
\ket{bc} & = \; \ket{b}\otimes\ket{c} & =\; \bigl[\;\0\;\;\0\;\;\1\;\;\1\;\bigr]^\mathrm{T}\;, \\
\ket{ca} & = \; \ket{c}\otimes\ket{a} & =\; \bigl[\;\1\;\;\0\;\;\1\;\;\0\;\bigr]^\mathrm{T}\;, \\
\ket{cb} & = \; \ket{c}\otimes\ket{b} & =\; \bigl[\;\0\;\;\1\;\;\0\;\;\1\;\bigr]^\mathrm{T}\;, \\
\ket{cc} & = \; \ket{c}\otimes\ket{c} & =\; \bigl[\;\1\;\;\1\;\;\1\;\;\1\;\bigr]^\mathrm{T}\;.
\end{array}
\end{equation}
The six entangled states can be classified according to their transformation properties
under global $S_3$ `rotations.'

The first is a singlet which transforms into itself under all permutations of $S_3$:
\begin{equation}
\bket{S}
\;=\; \ket{aa}+\ket{bb}+\ket{cc}
\;=\; \bigl[\;\0\;\;\1\;\;\1\;\;\0\;\bigr]^\mathrm{T} \;.
\end{equation}
This state is the closest analog to the spin singlet state
$\ket{0,0}=\frac{1}{\sqrt{2}}\left( \ket{\!\uparrow\downarrow}-\ket{\!\downarrow\uparrow} \right)$
in canonical QM,
as can be seen from the fact that $\ket{S}$ can also be written as
\begin{eqnarray}
\bket{S}
\;=\; \ket{ab}+\ket{ba}
\;=\; \ket{bc}+\ket{cb}
\;=\; \ket{ca}+\ket{ac}
\;.
\end{eqnarray}
This state is symmetric, however, under the interchange of the two particles since there is no
analog of $-1$ in $\mathbb{Z}_2$.

Three more states are symmetric, and transform as a triplet:
\begin{equation}
\begin{array}{lll}
\bket{(ab)} 
& =\; \ket{ab}+\ket{ba}+\ket{cc}
& =\; \bigl[\;\1\;\;\0\;\;\0\;\;\1\;\bigr]^\mathrm{T} \;,\\
\bket{(bc)}
& =\; \ket{aa}+\ket{bc}+\ket{cb}
& =\; \bigl[\;\1\;\;\1\;\;\1\;\;\0\;\bigr]^\mathrm{T} \;,\\
\bket{(ca)}
& =\; \ket{ac}+\ket{bb}+\ket{ca}
& =\; \bigl[\;\0\;\;\1\;\;\1\;\;\1\;\bigr]^\mathrm{T} \;.\\
\end{array}
\end{equation}
These would be the analog of the spin-one triplet in canonical QM.

The remaining two states are asymmetric under the interchange of the two particles:
\begin{equation}
\begin{array}{lll}
\bket{(abc)}
& =\; \ket{ab}+\ket{bc}+\ket{ca}
& =\; \bigl[\;\1\;\;\1\;\;\0\;\;\1\;\bigr]^\mathrm{T} \;,\\
\bket{(acb)}
& =\; \ket{ac}+\ket{cb}+\ket{ba}
& =\; \bigl[\;\1\;\;\0\;\;\1\;\;\1\;\bigr]^\mathrm{T} \;.\\
\end{array}
\end{equation}
These transform as a doublet: they are invariant under even permutations, but transform into each 
other under odd permtations.
There is a one-to-one correspondence between these states and the elements of $S_3$,
as well as a correspondence between the state multiplets and the conjugate classes of $S_3$.

\subsection{Geometric Characterization}

As discussed in section~II, the space of the fifteen state vectors in 
$GQM(4,2)$ possesses the projective geometry $PG(3,2)$.
In this geometry, three `points' $\ket{r}$, $\ket{s}$, $\ket{t}$ are on a `line' if
they add up to the zero vector.  The 15 `points' lie on 35 `lines,' with 7 `lines' crossing at each `point.'
The 35 `lines' are contained in 15 `planes,' with 3 `planes' intersecting at each `line.'

In the current context, the nine product states are `points' that 
lie on six `lines,' no three of which are in the same `plane,'
forming a `non-planar' grid: 
\begin{equation}
\begin{array}{ccccc}
\ket{aa} & =\!= & \ket{ab} & =\!= & \ket{ac} \\
\Vert & & \Vert & & \Vert \\
\ket{ba} & =\!= & \ket{bb} & =\!= & \ket{bc} \\
\Vert & & \Vert & & \Vert \\
\ket{ca} & =\!= & \ket{cb} & =\!= & \ket{cc}
\end{array}
\end{equation}
The six entangled states are each a sum of three product state `points,' no two of which lie on the 
same `line' of this grid, that is, no two states in the sum share the same `row' or `column.'

Beyond this, we have been unsuccessful in finding a geometric characterization, or
differentiation, of product and entangled states.
It is unclear whether a similar characterization is possible in cases other than $q=2$.
The discovery of a geometrical understanding applicable to the generic $GQM(4,q)$ case
with $PG(3,q)$ geometry could be enlightening.

\subsection{Local Rotations}

It will be useful to see how the six entangled states listed above transform 
into each other under `rotations' of either the first, or the second particle only.
We find:
\begin{equation}
\begin{array}{rlrl}
(ab)_1\ket{S}  & = & (ab)_2\ket{S}  & =\; \ket{(ab)}\;,\\
(bc)_1\ket{S}  & = & (bc)_2\ket{S}  & =\; \ket{(bc)}\;,\\
(ca)_1\ket{S}  & = & (ca)_2\ket{S}  & =\; \ket{(ca)}\;,\\
(acb)_1\ket{S} & = & (abc)_2\ket{S} & =\; \ket{(abc)}\;,\\
(abc)_1\ket{S} & = & (acb)_2\ket{S} & =\; \ket{(acb)}\;.
\end{array}
\label{ketPermutations}
\end{equation}
The transformation properties of the other states can be obtained from these
relations, for instance:
\begin{equation}
(ab)_1\ket{(bc)}\;=\;(ab)_1(bc)_1\ket{S}\;=\;(abc)_1\ket{S}\;=\;\ket{(acb)}\;.
\end{equation}
The fact that all six entangled states transform into each other this way
means that they are all equivalent and equally entangled, since the transformations considered
here amount to simple relabelings of the states in the two vector spaces that
are tensored.

\subsection{Two Particle Observables}

There are fifteen non-zero dual vectors in $V_2^*\otimes V_2^*$, but we will only be
looking at the nine product observables constructed from the nine product dual vectors
which are of the form
\begin{equation}
A_{rs}A_{tu}
\,=\,\{
\,\bra{\bar{r}}\otimes\bra{\bar{t}}\,,
\,\bra{\bar{r}}\otimes\bra{\bar{u}}\,,
\,\bra{\bar{s}}\otimes\bra{\bar{t}}\,,
\,\bra{\bar{s}}\otimes\bra{\bar{u}}\,
\}\;,
\end{equation}
where the indices $rs$ and $tu$ are $ab$, $bc$, or $ca$.
The four tensored dual vectors in this expression respectively represent the outcomes,
$++$, $+-$, $-+$, and $--$ when $A_{rs}A_{tu}$ is measured.
The row vector representations of all nine tensor products are given by
\begin{eqnarray}
\bra{\bar{a}}\otimes\bra{\bar{a}}
& = & \bigl[\,\0\;\;\0\;\;\0\;\;\1\,\bigr] \;,\cr
\bra{\bar{a}}\otimes\bra{\bar{b}}
& = & \bigl[\,\0\;\;\0\;\;\1\;\;\0\,\bigr] \;,\cr
\bra{\bar{a}}\otimes\bra{\bar{c}}
& = & \bigl[\,\0\;\;\0\;\;\1\;\;\1\,\bigr] \;,\cr
\bra{\bar{b}}\otimes\bra{\bar{a}}
& = & \bigl[\,\0\;\;\1\;\;\0\;\;\0\,\bigr] \;,\cr
\bra{\bar{b}}\otimes\bra{\bar{b}}
& = & \bigl[\,\1\;\;\0\;\;\0\;\;\0\,\bigr] \;,\cr
\bra{\bar{b}}\otimes\bra{\bar{c}}
& = & \bigl[\,\1\;\;\1\;\;\0\;\;\0\,\bigr] \;,\cr
\bra{\bar{c}}\otimes\bra{\bar{a}}
& = & \bigl[\,\0\;\;\1\;\;\0\;\;\1\,\bigr] \;,\cr
\bra{\bar{c}}\otimes\bra{\bar{b}}
& = & \bigl[\,\1\;\;\0\;\;\1\;\;\0\,\bigr] \;,\cr
\bra{\bar{c}}\otimes\bra{\bar{c}}
& = & \bigl[\,\1\;\;\1\;\;\1\;\;\1\,\bigr] \;.
\end{eqnarray}
%

\subsection{Probabilities and Correlations}

Applying \eref{Pdef} to product observables,
the probability of obtaining an outcome $(x,y)$ represented by the
product dual vector $\bra{xy}=\bra{x}\otimes\bra{y}$ when observable $O_1O_2$ is
measured on state $\ket{\psi}$ is given by:
\begin{equation}
P(O_1O_2\,;xy\,|\,\psi) \;=\; 
\frac{|\braket{xy}{\psi}|^2}
{\sum_{zw}|\braket{zw}{\psi}|^2}
\;.
\end{equation}
For a product state $\ket{\psi}=\ket{r}\otimes\ket{s}\equiv\ket{rs}$, the brackets factorize:
\begin{equation}
\braket{xy}{rs}
\;=\; \bigl(\bra{x}\otimes\bra{y}\,\bigr)\bigl(\,\ket{r}\otimes\ket{s}\bigr)
\;=\; \braket{x}{r}\braket{y}{s}\;,
\end{equation}
and due to the condition \eref{ProductPreservingMap} we imposed on the absolute values,
we have
\begin{equation}
\bigl|\braket{xy}{rs}\bigr|\;=\;\bigl|\braket{x}{r}\bigr|\,\bigl|\braket{y}{s}\bigr|\;.
\end{equation}
Consequently, the probability also factorizes as
\begin{eqnarray}
P(O_1O_2\,;xy\,|\,rs) 
& = & 
\frac{|\braket{xy}{rs}|^2}
{\sum_{zw}|\braket{zw}{rs}|^2}
\cr
& = &
\frac{|\braket{x}{r}|^2|\braket{y}{s}|^2}
{\sum_{z}\sum_{w}
|\braket{z}{r}|^2|\braket{w}{s}|^2}
\cr
& = & 
\left[
\frac{|\braket{x}{r}|^2}
{\sum_{z}|\braket{z}{r}|^2}
\right]
\left[
\frac{\braket{y}{s}|^2}
{\sum_{w}|\braket{w}{s}|^2}
\right]
\cr
& = &
P(O_1\,;x\,|\,r)\,P(O_2\,;y\,|\,s)\;,
\phantom{\frac{X}{X}}
\end{eqnarray}
which is a property we would like to preserve for unentangled states.
Otherwise, no isolated particle would be possible.
Note the importance of \eref{ProductPreservingMap} for this factorization to occur.
The expectation value of the product observable $O_1O_2$, \textit{i.e.} the
correlation between $O_1$ and $O_2$, will be
\begin{equation}
\vev{O_1O_2}_\psi 
\;=\;
\sum_{xy}\,xy\;P(O_1O_2\,;xy\,|\,\psi)
\;=\;
\frac{\sum_{xy} xy\,|\braket{xy}{\psi}|^2}
{\sum_{zw}|\braket{zw}{\psi}|^2}
\;,
\end{equation}
which for product states factorizes as
\begin{equation}
\vev{O_1O_2}_{rs} \;=\; \vev{O_1}_r\vev{O_2}_s\;.
\end{equation}
Thus, the correlations for the nine product states
will simply be products of those listed in \tref{oneparticlevevs}.

The probabilities and expectation values of the nine product observables
for all six entangled states are shown in \tref{twoparticlevevs},
where we have used the notation $X=A_{bc}$, $Y=A_{ca}$, $Z=A_{ab}$.
The entries can be `rotated' into each other 
via \eref{braPermutations} and \eref{ketPermutations}.
For instance, since 
$(ab)_1 X_1 X_2 \;=\; -Y_1 X_2$ and $(ab)_1\ket{(ab)}=\ket{S}$, 
we have
\begin{eqnarray}
P(Y_1X_2;++,S)
& = & P(-Y_1 X_2;-+,S)  \cr
& = & (ab)_1 P(X_1 X_2;-+,(ab)) 
\;=\; \frac{1}{3}\;,\qquad
\end{eqnarray} 
and so on.

\begin{landscape}

\fulltable{\label{twoparticlevevs}Correlations of observables for the six entangled states.\\}
\scalebox{0.88}{
\begin{tabular}{ccccccc}
\br
& state & $++$ & $+-$ & $-+$ & $--$ & E.V. \\
\br
$X_1X_2$ 
& $S$ & $0$ & $\frac{1}{2}$ & $\frac{1}{2}$ & $0$ & $-1$ \phantom{\Big|}\\
\cline{2-7}
& $(ab)$ & $\frac{1}{3}$ & $\frac{1}{3}$ & $\frac{1}{3}$ & $0$ & $-\frac{1}{3}$ \phantom{\Big|}\\
\cline{2-7}
& $(bc)$ & $\frac{1}{2}$ & $0$ & $0$ & $\frac{1}{2}$ & $+1$ \phantom{\Big|}\phantom{\Big|}\\
\cline{2-7}
& $(ca)$ & $0$ & $\frac{1}{3}$ & $\frac{1}{3}$ & $\frac{1}{3}$ & $-\frac{1}{3}$ \phantom{\Big|}\\
\cline{2-7}
& $(abc)$ & $\frac{1}{3}$ & $0$ & $\frac{1}{3}$ & $\frac{1}{3}$ & $+\frac{1}{3}$ \phantom{\Big|}\\
\cline{2-7}
& $(acb)$ & $\frac{1}{3}$ & $\frac{1}{3}$ & $0$ & $\frac{1}{3}$ & $+\frac{1}{3}$ \phantom{\Big|}\\
\br
$X_1Y_2$
& $S$ & $\frac{1}{3}$ & $\frac{1}{3}$ & $0$ & $\frac{1}{3}$ & $+\frac{1}{3}$ \phantom{\Big|}\\
\cline{2-7}
& $(ab)$ & $\frac{1}{2}$ & $0$ & $0$ & $\frac{1}{2}$ & $+1$ \phantom{\Big|}\\
\cline{2-7}
& $(bc)$ & $0$ & $\frac{1}{3}$ & $\frac{1}{3}$ & $\frac{1}{3}$ & $-\frac{1}{3}$ \phantom{\Big|}\\
\cline{2-7}
& $(ca)$ & $\frac{1}{3}$ & $\frac{1}{3}$ & $\frac{1}{3}$ & $0$ & $-\frac{1}{3}$ \phantom{\Big|}\\
\cline{2-7}
& $(abc)$ & $0$ & $\frac{1}{2}$ & $\frac{1}{2}$ & $0$ & $-1$ \phantom{\Big|}\\
\cline{2-7}
& $(acb)$ & $\frac{1}{3}$ & $0$ & $\frac{1}{3}$ & $\frac{1}{3}$ & $+\frac{1}{3}$ \phantom{\Big|}\\
\br
$X_1Z_2$
& $S$ & $\frac{1}{3}$ & $0$ & $\frac{1}{3}$ & $\frac{1}{3}$ & $+\frac{1}{3}$ \phantom{\Big|}\\
\cline{2-7}
& $(ab)$ & $0$ & $\frac{1}{3}$ & $\frac{1}{3}$ & $\frac{1}{3}$ & $-\frac{1}{3}$ \phantom{\Big|}\\
\cline{2-7}
& $(bc)$ & $\frac{1}{3}$ & $\frac{1}{3}$ & $\frac{1}{3}$ & $0$ & $-\frac{1}{3}$ \phantom{\Big|}\\
\cline{2-7}
& $(ca)$ & $\frac{1}{2}$ & $0$ & $0$ & $\frac{1}{2}$ & $+1$ \phantom{\Big|}\\
\cline{2-7}
& $(abc)$ & $\frac{1}{3}$ & $\frac{1}{3}$ & $0$ & $\frac{1}{3}$ & $+\frac{1}{3}$ \phantom{\Big|}\\
\cline{2-7}
& $(acb)$ & $0$ & $\frac{1}{2}$ & $\frac{1}{2}$ & $0$ & $-1$ \phantom{\Big|}\\
\br
\end{tabular}
\hspace{0.1cm}
\begin{tabular}{ccccccc}
\br
& state & $++$ & $+-$ & $-+$ & $--$ & E.V. \\
\br
$Y_1X_2$ 
& $S$ & $\frac{1}{3}$ & $0$ & $\frac{1}{3}$ & $\frac{1}{3}$ & $+\frac{1}{3}$ \phantom{\Big|}\\
\cline{2-7}
& $(ab)$ & $\frac{1}{2}$ & $0$ & $0$ & $\frac{1}{2}$ & $+1$ \phantom{\Big|}\\
\cline{2-7}
& $(bc)$ & $0$ & $\frac{1}{3}$ & $\frac{1}{3}$ & $\frac{1}{3}$ & $-\frac{1}{3}$ \phantom{\Big|}\\
\cline{2-7}
& $(ca)$ & $\frac{1}{3}$ & $\frac{1}{3}$ & $\frac{1}{3}$ & $0$ & $-\frac{1}{3}$ \phantom{\Big|}\\
\cline{2-7}
& $(abc)$ & $\frac{1}{3}$ & $\frac{1}{3}$ & $0$ & $\frac{1}{3}$ & $+\frac{1}{3}$ \phantom{\Big|}\\
\cline{2-7}
& $(acb)$ & $0$ & $\frac{1}{2}$ & $\frac{1}{2}$ & $0$ & $-1$ \phantom{\Big|}\\
\br
$Y_1Y_2$
& $S$ & $0$ & $\frac{1}{2}$ & $\frac{1}{2}$ & $0$ & $-1$ \phantom{\Big|}\\
\cline{2-7}
& $(ab)$ & $0$ & $\frac{1}{3}$ & $\frac{1}{3}$ & $\frac{1}{3}$ & $-\frac{1}{3}$ \phantom{\Big|}\\
\cline{2-7}
& $(bc)$ & $\frac{1}{3}$ & $\frac{1}{3}$ & $\frac{1}{3}$ & $0$ & $-\frac{1}{3}$ \phantom{\Big|}\\
\cline{2-7}
& $(ca)$ & $\frac{1}{2}$ & $0$ & $0$ & $\frac{1}{2}$ & $+1$ \phantom{\Big|}\\
\cline{2-7}
& $(abc)$ & $\frac{1}{3}$ & $0$ & $\frac{1}{3}$ & $\frac{1}{3}$ & $+\frac{1}{3}$ \phantom{\Big|}\\
\cline{2-7}
& $(acb)$ & $\frac{1}{3}$ & $\frac{1}{3}$ & $0$ & $\frac{1}{3}$ & $+\frac{1}{3}$ \phantom{\Big|}\\
\br
$Y_1Z_2$
& $S$ & $\frac{1}{3}$ & $\frac{1}{3}$ & $0$ & $\frac{1}{3}$ & $+\frac{1}{3}$ \phantom{\Big|}\\
\cline{2-7}
& $(ab)$ & $\frac{1}{3}$ & $\frac{1}{3}$ & $\frac{1}{3}$ & $0$ & $-\frac{1}{3}$ \phantom{\Big|}\\
\cline{2-7}
& $(bc)$ & $\frac{1}{2}$ & $0$ & $0$ & $\frac{1}{2}$ & $+1$ \phantom{\Big|}\\
\cline{2-7}
& $(ca)$ & $0$ & $\frac{1}{3}$ & $\frac{1}{3}$ & $\frac{1}{3}$ & $-\frac{1}{3}$ \phantom{\Big|}\\
\cline{2-7}
& $(abc)$ & $0$ & $\frac{1}{2}$ & $\frac{1}{2}$ & $0$ & $-1$ \phantom{\Big|}\\
\cline{2-7}
& $(acb)$ & $\frac{1}{3}$ & $0$ & $\frac{1}{3}$ & $\frac{1}{3}$ & $+\frac{1}{3}$ \phantom{\Big|}\\
\br
\end{tabular}
\hspace{0.1cm}
\begin{tabular}{ccccccc}
\br
& state & $++$ & $+-$ & $-+$ & $--$ & E.V. \\
\br
$Z_1X_2$ 
& $S$ & $\frac{1}{3}$ & $\frac{1}{3}$ & $0$ & $\frac{1}{3}$ & $+\frac{1}{3}$ \phantom{\Big|}\\
\cline{2-7}
& $(ab)$ & $0$ & $\frac{1}{3}$ & $\frac{1}{3}$ & $\frac{1}{3}$ & $-\frac{1}{3}$ \phantom{\Big|}\\
\cline{2-7}
& $(bc)$ & $\frac{1}{3}$ & $\frac{1}{3}$ & $\frac{1}{3}$ & $0$ & $-\frac{1}{3}$ \phantom{\Big|}\\
\cline{2-7}
& $(ca)$ & $\frac{1}{2}$ & $0$ & $0$ & $\frac{1}{2}$ & $+1$ \phantom{\Big|}\\
\cline{2-7}
& $(abc)$ & $0$ & $\frac{1}{2}$ & $\frac{1}{2}$ & $0$ & $-1$ \phantom{\Big|}\\
\cline{2-7}
& $(acb)$ & $\frac{1}{3}$ & $0$ & $\frac{1}{3}$ & $\frac{1}{3}$ & $+\frac{1}{3}$ \phantom{\Big|}\\
\br
$Z_1Y_2$
& $S$ & $\frac{1}{3}$ & $0$ & $\frac{1}{3}$ & $\frac{1}{3}$ & $+\frac{1}{3}$ \phantom{\Big|}\\
\cline{2-7}
& $(ab)$ & $\frac{1}{3}$ & $\frac{1}{3}$ & $\frac{1}{3}$ & $0$ & $-\frac{1}{3}$ \phantom{\Big|}\\
\cline{2-7}
& $(bc)$ & $\frac{1}{2}$ & $0$ & $0$ & $\frac{1}{2}$ & $+1$ \phantom{\Big|}\\
\cline{2-7}
& $(ca)$ & $0$ & $\frac{1}{3}$ & $\frac{1}{3}$ & $\frac{1}{3}$ & $-\frac{1}{3}$ \phantom{\Big|}\\
\cline{2-7}
& $(abc)$ & $\frac{1}{3}$ & $\frac{1}{3}$ & $0$ & $\frac{1}{3}$ & $+\frac{1}{3}$ \phantom{\Big|}\\
\cline{2-7}
& $(acb)$ & $0$ & $\frac{1}{2}$ & $\frac{1}{2}$ & $0$ & $-1$ \phantom{\Big|}\\
\br
$Z_1Z_2$
& $S$ & $0$ & $\frac{1}{2}$ & $\frac{1}{2}$ & $0$ & $-1$ \phantom{\Big|}\\
\cline{2-7}
& $(ab)$ & $\frac{1}{2}$ & $0$ & $0$ & $\frac{1}{2}$ & $+1$ \phantom{\Big|}\\
\cline{2-7}
& $(bc)$ & $0$ & $\frac{1}{3}$ & $\frac{1}{3}$ & $\frac{1}{3}$ & $-\frac{1}{3}$ \phantom{\Big|}\\
\cline{2-7}
& $(ca)$ & $\frac{1}{3}$ & $\frac{1}{3}$ & $\frac{1}{3}$ & $0$ & $-\frac{1}{3}$ \phantom{\Big|}\\
\cline{2-7}
& $(abc)$ & $\frac{1}{3}$ & $0$ & $\frac{1}{3}$ & $\frac{1}{3}$ & $+\frac{1}{3}$ \phantom{\Big|}\\
\cline{2-7}
& $(acb)$ & $\frac{1}{3}$ & $\frac{1}{3}$ & $0$ & $\frac{1}{3}$ & $+\frac{1}{3}$ \phantom{\Big|}\\
\br
\end{tabular}
}
\endfulltable

\end{landscape}

\subsection{Entanglement and the Impossibility of Hidden Variables}

\begin{figure}[t]
\caption{\label{Implications}The implication chart for the state $\ket{S}$.
Arrows point from the condition toward the implication.
By tracing the arrows, it is easy to see that
no classical configurations, and thus no hidden variable theory,
can satisfy all of these requirements.
If we ignore the observable $X$ and look at only $Y$ and $Z$, then
the assignments within the dashed boxes are possible classical configurations.
However, neither allow for the pairs $(Y_1 Z_2)$ and $(Z_1 Y_2)$ to be
anti-correlated, which occurs with probability $1/3$ in our QM.\\
}
\begin{indented}\item[]
\includegraphics[width=10cm]{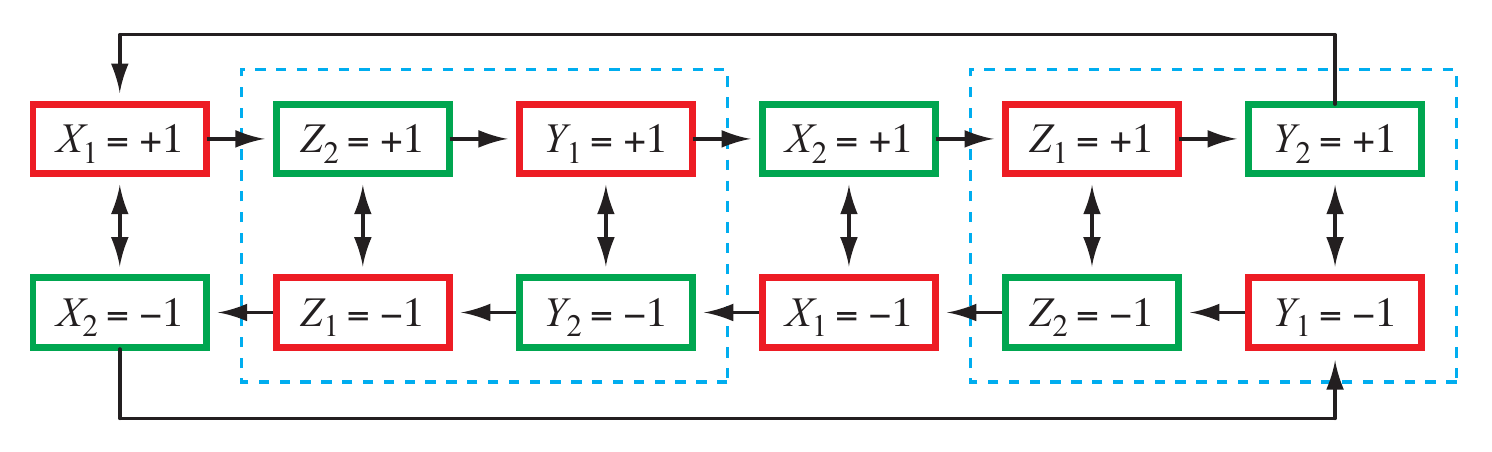}
\end{indented}
\end{figure}

We now demonstrate that hidden variables cannot reproduce the probabilities and
correlations predicted by GQM for entangled states.
The argument is analogous to those of Greenberger, Horne, Shimony, and Zeilinger \cite{GHZ},
and of Hardy \cite{Hardy:1993zza} for canonical QM.

Since all six entangled states are equivalent, it suffices to consider
only one, for which we will use the state $\ket{S}$.
From \tref{twoparticlevevs}, we see that
\begin{equation}
\begin{array}{lll}
P(X_1X_2;++,S) & =\;
P(X_1X_2;--,S) & =\; 0\;, \\
P(\,Y_1\,Y_2\,;++,S) & =\; 
P(\,Y_1\,Y_2\,;--,S) & =\; 0\;, \\
P(\,Z_1Z_2\,;++,S) & =\; 
P(\,Z_1Z_2\,;--,S) & =\; 0\;, 
\end{array}
\end{equation}
which means that the pairs $(X_1,X_2)$, $(Y_1,Y_2)$, and $(Z_1,Z_2)$ are all 
completely anti-correlated.
We next note that
\begin{equation}
P(X_1Z_2;+-,S) \;=\; 0\;,
\end{equation}
which means that $X_1=+1$ necessarily implies $Z_2=+1$,
while $Z_2=-1$ necessarily implies $X_1=-1$.  
Similarly,
\begin{equation}
P(Y_1Z_2;-+,S) \;=\; 0\;
\end{equation}
means that $Z_2=+1$ necessarily implies $Y_1=+1$,
while $Y_1=-1$ necessarily implies $Z_2=-1$.  
Going through \tref{twoparticlevevs} in this fashion,
we obtain the implication diagram shown in \fref{Implications}.
As is clear from the diagram, no classical configuration exists which would be
compatible with all of these constraints.
For instance, $X_1=+1$ implies $Z_2=+1$, which implies $Y_1=+1$,
which implies $X_2=+1$, which contradicts the requirement that
$X_1$ and $X_2$ are anti-correlated.

It should be noted that even if we limit our attention to only two of the three
observables available for each particle, hidden variables still cannot reproduce
the quantum probabilities.
For instance, consider only $Y$ and $Z$ for both particles.  Then,
the selection of values within the dashed boxes on \fref{Implications} 
give possible classical configurations. However, the combinations
$(Y_1 Z_2)=(+-)$ and $(Z_1 Y_2)=(-+)$ cannot occur even though they are
possible quantum mechanically.  
Thus, the entangled states in our model are truly `quantum' 
in the sense discussed in the introduction, and entangled.

\subsection{The CHSH Bound}

Let us now find the CHSH bound of our model, \textit{i.e.} the upper bound of the
absolute value of the CHSH correlator defined in \eref{CHSHcorrdef}.
Owing to the equivalence of all entangled states, we only need to
look at the correlations for one state for all possible observable combinations.
That is, using \eref{braPermutations} and \eref{ketPermutations}, we can 
convert the correlations for any entangled state into those for the state $\ket{S}$.
For instance:
\begin{eqnarray}
\vev{X_1,Y_1;X_2,Y_2}_{(ab)}
& = & -\vev{Y_1,X_1;X_2,Y_2}_{S} \cr
& = & -\vev{X_1,Y_1;Y_2,X_2}_{S} \;.
\end{eqnarray}
We also need not consider the negatives of the observables 
as long as all possible choices for $A_1$, $A_2$, $B_1$, and $B_2$ are considered since
\begin{eqnarray}
\lefteqn{\vev{A_1,A_2\,;B_1,B_2}} \cr
& = & \vev{A_1,-A_2\,;B_2,B_1}
\;=\; -\vev{-A_1,A_2\,;B_2,B_1} \cr
& = & \vev{A_2,A_1\,;B_1,-B_2}
\;=\; -\vev{A_2,A_1\,;-B_1,B_2}\;.
\end{eqnarray}
Then, from simple inspection of \tref{twoparticlevevs}, we can see that
the maximum absolute value of the CHSH correlator is achieved for
\begin{eqnarray}
\vev{X,Y;Y,X}_S 
& = & -2\;,\cr
\vev{X,Z;Y,Z}_S 
& = & +2\;,
\end{eqnarray}
with arbitrary permutations of the three observables leading to the same values.
All the other correlators yield $\pm \frac{2}{3}$.
Thus, the CHSH bound for our model is 2, the classical value,
despite the fact that it is `quantum' and does not allow for any hidden variables.

That this bound is different from the Cirel'son bound of $2\sqrt{2}$ for spin systems in canonical QM should not be a surprise \cite{cirelson}.   The underlying vector space in the present `spin' system has no inner product, a key element in the derivation of the Cirel'son bound \cite{Chang:2011yt}, and the symmetry group in this context is not equivalent to $SU(2)$.   The entangled states have parallels to those of $SU(2)$, but have an independent existence and structure.   These features conspire to lower the bound from the canonical value.   
While its agreement with the classical value seems to be a happenstance,
it demonstrates that the CHSH bound by itself does not necessarily distinguish between
quantum and classical descriptions.


\section{`Rotations' and Rotations}

We now consider the cases $q=3$, $4$, and $5$.
Since the details of how $GQM(2,q)$ is implemented for these cases, and the calculation of the
CHSH bound are not that different from the $q=2$ case, we will not go into detail about these
cases here and refer the reader to the argument presented in \cite{Chang:2012} which
applies to all values of $q$.
In this section, we will mostly be concerned with how the `spin directions' in $GQM(2,q)$ can be
mapped to actual directions in 3D space, and how the elements of $PGL(2,q)$ can be identified
with $SO(3)$ rotations.

\subsection{$\mathbb{Z}_3$ case}

$GQM(2,3)$ is constructed on the field consisting of three elements,
$GF(3)=\mathbb{Z}_3=\mathbb{Z}/3\mathbb{Z}=\{\0,\1,\2\}$, with
addition and multiplication tables given by
\begin{center}
\begin{tabular}[t]{c|lll}
$\;+\;$ & $\;\0\;$ & $\;\1\;$ & $\;\2\;$ \\
\hline
$\0$    & $\;\0$   & $\;\1$   & $\;\2$   \\
$\1$    & $\;\1$   & $\;\2$   & $\;\0$   \\
$\2$    & $\;\2$   & $\;\0$   & $\;\1$  
\end{tabular}
\qquad
\begin{tabular}[t]{c|lll}
$\;\times\;$ & $\;\0\;$ & $\;\1\;$ & $\;\2\;$ \\
\hline
$\0$    & $\;\0$ & $\;\0$    & $\;\0$  \\
$\1$    & $\;\0$ & $\;\1$    & $\;\2$  \\
$\2$    & $\;\0$ & $\;\2$    & $\;\1$ 
\end{tabular}
\end{center}
In the following, we will write $\2$ as $-\1$.
Since $\mathbb{Z}_3\backslash\{\0\}=\{\1,-\1\}$, 
each physical state will be represented by two vectors in $V_3=\mathbb{Z}_3^2$
which differ by the multiplicative `phase' $-\1$.

Thus, of the $3^2-1=8$ non-zero vectors in $V_3$, there are pairs of
vectors that are equivalent, and the inequivalent ones can be taken to be:
\begin{equation}
\ket{a} = \left[\begin{array}{c}  \1 \\ \0 \end{array}\right],\;\;
\ket{b} = \left[\begin{array}{c}  \0 \\ \1 \end{array}\right],\;\;
\ket{c} = \left[\begin{array}{r} -\1 \\ \1 \end{array}\right],\;\;
\ket{d} = \left[\begin{array}{c}  \1 \\ \1 \end{array}\right].
\end{equation}
Note that any pair of these states can be written as the sum and difference of
the other two up to phases, \textit{e.g.}
\begin{equation}
\ket{c} \;=\; -\ket{a}+\ket{b}\;,\qquad
\ket{d} \;=\; \ket{a}+\ket{b}\;.
\end{equation}
%
Thus, a basis transformation which interchanges a pair of states
would leave the other two unaffected.
In the above example, interchanging $\ket{a}$ and $\ket{b}$
would leave $\ket{d}$ unchanged, while $\ket{c}$ only acquires an unphysical 
phase $-\1$.  Thus, single transpositions of the vector labels are possible,
and the group generated by those transpositions would be $S_4\cong PGL(2,3)$.
That is, all permutations of the vector labels are possible under
basis transformations.  This is the group of `rotations' for $GQM(2,3)$.

The inequivalent dual-vectors of $V_3^*$ can be taken to be
\begin{eqnarray}
\bra{\bar{a}} & = & \bigl[\;\0\;-\!\1\;\bigr]\;,\quad
\bra{\bar{b}} \;=\; \bigl[\;\1\;\quad\0\;\bigr]\;,\cr
\bra{\bar{c}} & = & \bigl[\;\1\;\quad\1\;\bigr]\;,\quad
\bra{\bar{d}} \;=\; \bigl[\;\1\;-\!\1\;\bigr]\;.
\end{eqnarray}
%
The actions of these dual-vectors on the vectors are given by:
\begin{center}
\begin{tabular}{|c||c|c|c|c|}
\hline
& $\;\;\ket{a}\;\;$ & $\phantom{-}\ket{b}\;\;$ & $\phantom{-}\ket{c}\;\;$ & $\phantom{-}\ket{d}\;\;$ \\
\hline
$\;\;\bra{\bar{a}}\;\;$ & $\0$ & $-\1$           & $-\1$           & $-\1$ \\
$\bra{\bar{b}}$         & $\1$ & $\phantom{-}\0$ & $-\1$           & $\phantom{-}\1$ \\
$\bra{\bar{c}}$         & $\1$ & $\phantom{-}\1$ & $\phantom{-}\0$ & $-\1$ \\
$\bra{\bar{d}}$         & $\1$ & $-\1$           & $\phantom{-}\1$ & $\phantom{-}\0$ \\
\hline
\end{tabular}
\end{center}
Thus,
\begin{eqnarray}
\braket{\bar{r}}{s} 
& =    & \0 \quad \mbox{if $r=s$,}     \cr
& \neq & \0 \quad \mbox{if $r\neq s$,}
\end{eqnarray}
and the relation $|\braket{\bar{r}}{s}|=1-\delta_{rs}$ is obtained
in this case also. Maintaining this relation would require
relabeling the dual vectors in the same way as the vectors
under basis transformations.

\begin{figure}[t]
\caption{\label{Z3spins}To map `rotations' in {$PGL(2,3)\cong S_4$} to rotations in {$SO(3)$}: (a) label the faces of an octahedron with four symbols as shown.  Then, every permutation of the four labels {$abcd$} will correspond a rotation of the octahedral group.  
(b) The `spin' observable {$A_{ab}$} in {$GQM(2,3)$} can be mapped onto a direction in 3D space as shown.
(c) The urchin diagram showing all 12 `spin' directions $GQM(2,3)$.\\}
\subfigure[]{\includegraphics[width=4cm]{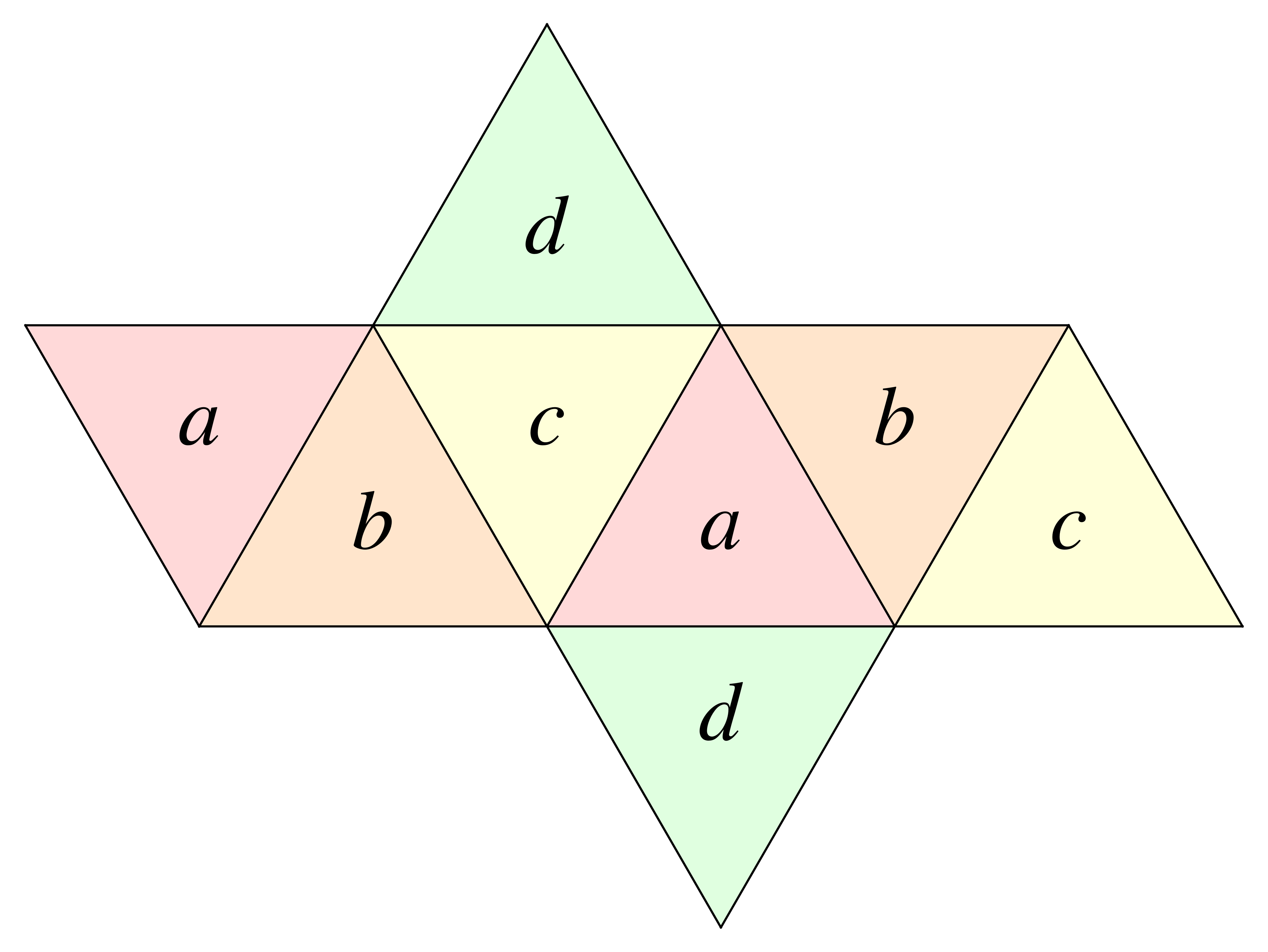}}
\subfigure[]{\includegraphics[height=3.7cm]{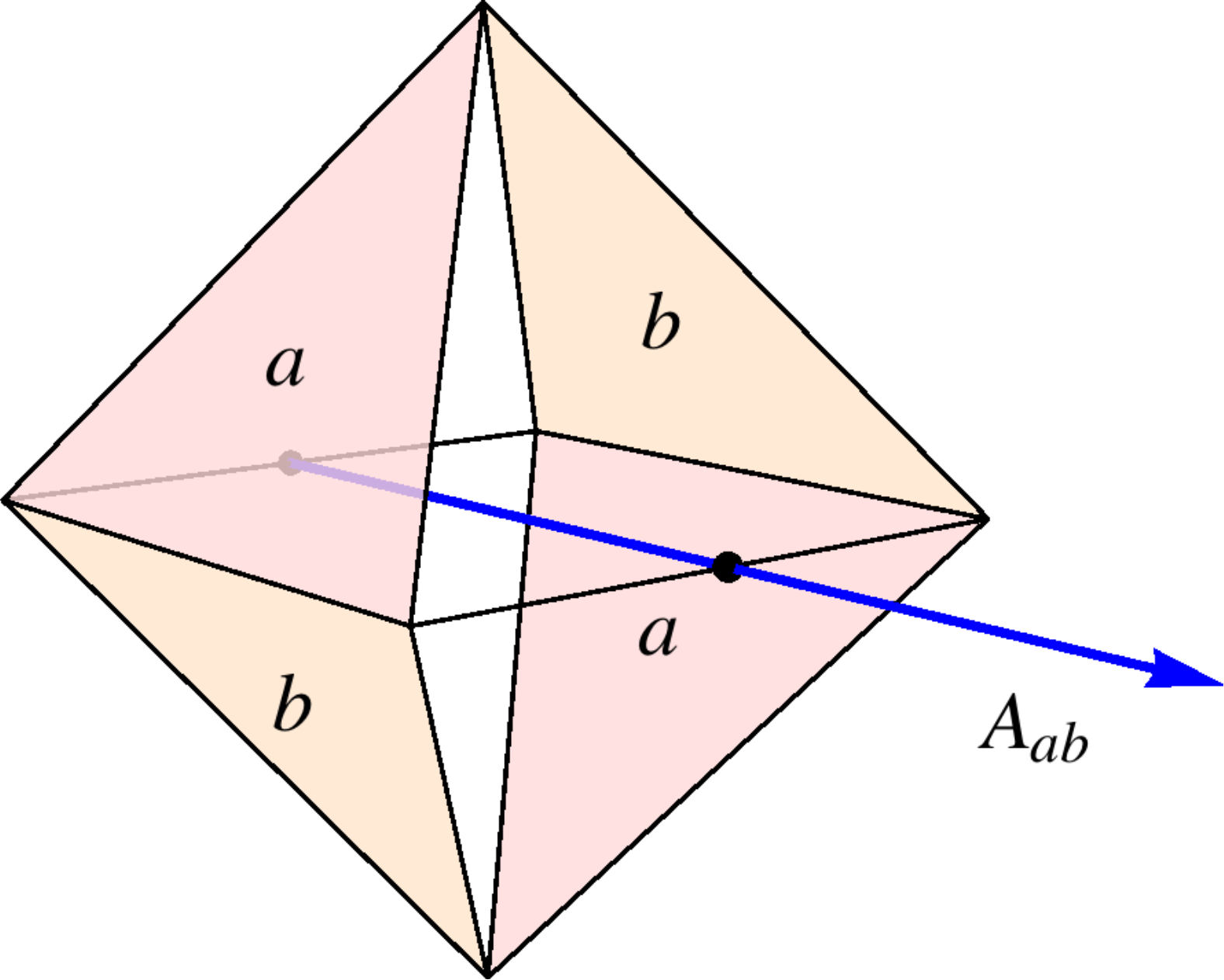}}
\subfigure[]{\includegraphics[height=3.7cm]{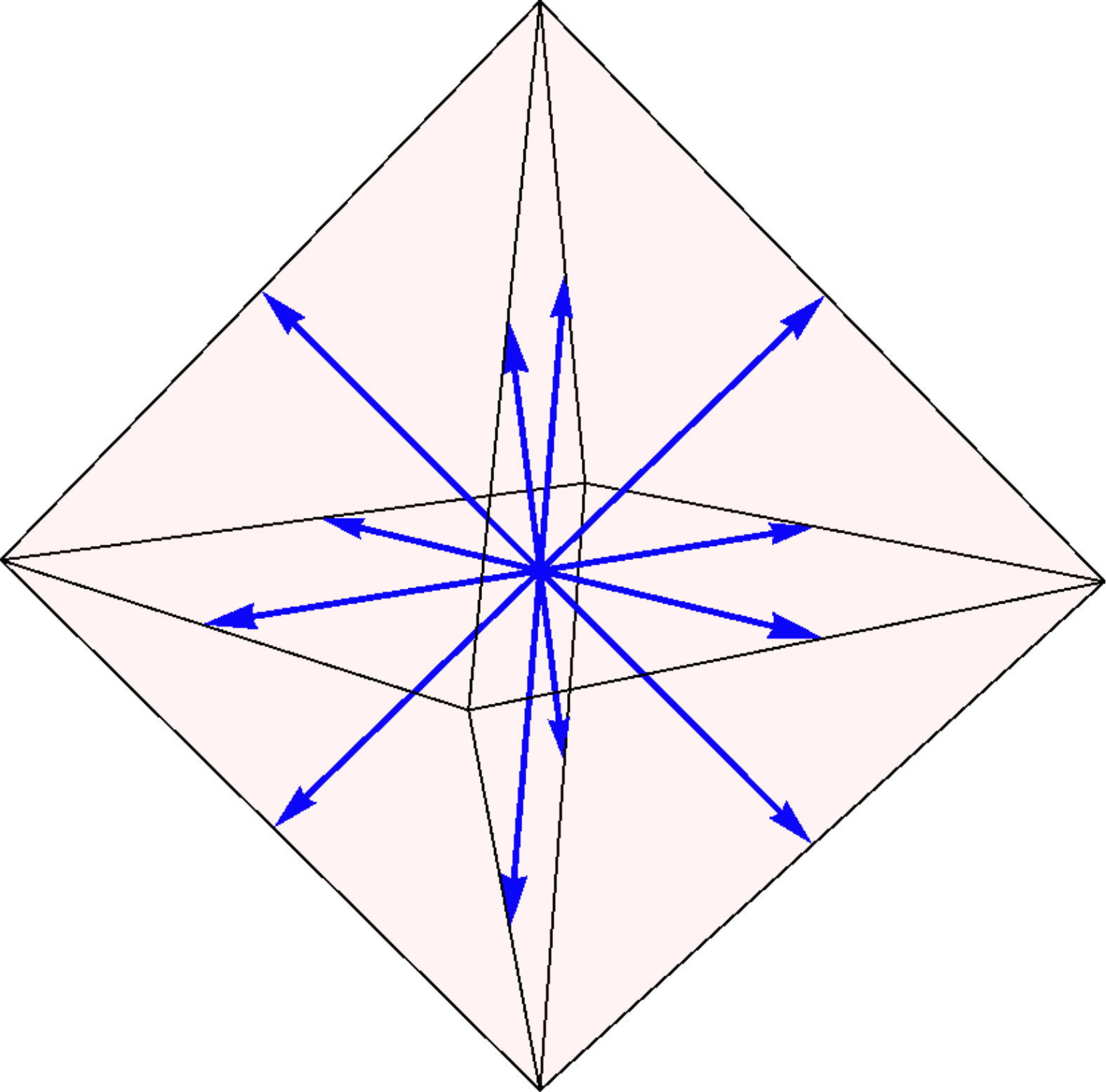}}
\end{figure}

From the set of four dual-vectors, we can define $4\times 3 = 12$ observables with outcomes
$\pm 1$, or 6 if we count the `spins' pointing in opposite directions as the same observable.
These `spins' can be associated with actual directions in 3D,
and their $PGL(2,3)\cong S_4$ transformations with rotations in $SO(3)$ as 
shown in \fref{Z3spins}: 
First, label the faces of an octahedron with four letters $abcd$, each 
letter appearing twice, on opposing faces as shown in \fref{Z3spins}(a).
The octahedral group $O$ \cite{Hamermesh} which rotates the octahedron onto itself consists of
24 elements.  Each of these elements will permute the four letters on the
faces of the octahedron.  Thus, there exists a one-to-one correspondence between
elements of the octahedral group and the 24 permutations of $S_4$.  The `direction' of the `spin' 
observable $A_{ab}$ can be associated with the direction of the arrow shown in \fref{Z3spins}(b).
All 12 `spin' directions can be mapped this way, and \fref{Z3spins}(c) shows the
resulting urchin of spin-directions.
These 12 `spins' transform into each other under $S_4\cong O$ rotations.

\subsection{$\mathbb{Z}_2[\uom]$ case}

$GQM(2,4)$ is constructed on the field consisting of four elements,
$GF(4)=\mathbb{Z}_2[\uom]=\{\0,\1,\uom,\uom^2\}$, 
which is the Galois extension of $\mathbb{Z}_2=\mathbb{Z}/2\mathbb{Z}$
with solutions to the equation
\begin{equation}
\underline{x}^2+\underline{x}+\1\;=\;\0\;,
\end{equation}
which we denote $\uom$ and $\uom^2=\1+\uom$.
The addition and multiplication tables of this field are given by
\begin{center}
\begin{tabular}[t]{l|llll}
$+\;$      & $\;\0\;$   & $\;\1\;$   & $\;\uom\;$ & $\;\uom^2\;$ \\
\hline
$\,\0$     & $\;\0$     & $\;\1$     & $\;\uom$   & $\;\uom^2$   \\
$\,\1$     & $\;\1$     & $\;\0$     & $\;\uom^2$ & $\;\uom$     \\
$\,\uom$   & $\;\uom$   & $\;\uom^2$ & $\;\0$     & $\;\1$       \\
$\,\uom^2$ & $\;\uom^2$ & $\;\uom$   & $\;\1$     & $\;\0$
\end{tabular}
\qquad
\begin{tabular}[t]{l|llll}
$\times\;$ & $\;\;\0\;$ & $\;\;\1\;$   & $\;\uom\;$ & $\;\uom^2\;$ \\
\hline
$\,\0$     & $\;\;\0$   & $\;\;\0$     & $\;\0$     & $\;\0$       \\
$\,\1$     & $\;\;\0$   & $\;\;\1$     & $\;\uom$   & $\;\uom^2$   \\
$\,\uom$   & $\;\;\0$   & $\;\;\uom$   & $\;\uom^2$ & $\;\1$       \\
$\,\uom^2$ & $\;\;\0$   & $\;\;\uom^2$ & $\;\1$     & $\;\uom$
\end{tabular}
\end{center}
Since $\mathbb{Z}_2[\uom]\backslash\{\0\}=\{\1,\uom,\uom^2\}$, 
each physical state will be represented by three vectors in $V_4=\{\,\mathbb{Z}_2[\uom]\,\}^2$
which differ by multiplicative `phases' $\uom$ or $\uom^2$.

Thus, of the $4^2-1=15$ non-zero vectors in $V_4$, every three of them are equivalent, 
and the $15/3=5$ inequivalent ones can be taken to be:
\begin{equation}
\begin{array}{lll}
\ket{a} = \left[\begin{array}{l} \1     \\ \0 \end{array}\right],\quad &
\ket{b} = \left[\begin{array}{l} \0     \\ \1 \end{array}\right],\quad &
\ket{c} = \left[\begin{array}{l} \uom   \\ \1 \end{array}\right],\quad \\
\ket{d} = \left[\begin{array}{l} \uom^2 \\ \1 \end{array}\right],\quad &
\ket{e} = \left[\begin{array}{l} \1     \\ \1 \end{array}\right],      & \\
\end{array}
\end{equation}
Let us choose a pair of vectors as a basis and express the other three as linear combinations
of those two, \textit{e.g.}
\begin{equation}
\ket{c} \,=\, \uom\,\ket{a}+\ket{b}\,,\quad
\ket{d} \,=\, \uom^2\ket{a}+\ket{b}\,,\quad
\ket{e} \,=\, \ket{a}+\ket{b}\,.
\end{equation}
%
Now consider a basis transformation that would interchange $\ket{a}$ and $\ket{b}$.
This would leave $\ket{e}$ unchanged, but $\ket{c}$ and $\ket{d}$ would
transform into each other:
\begin{eqnarray}
\ket{c} & \rightarrow & \ket{a}+\uom\,\ket{b}\;\cong\;\uom^2\ket{a}+\ket{b}\;=\;\ket{d}\;,\cr
\ket{d} & \rightarrow & \ket{a}+\uom^2\ket{b}\;\cong\;\uom\,\ket{a}+\ket{b}\;=\;\ket{c}\;.
\end{eqnarray}
Thus, single transpositions of the vector labels are impossible.
Transpositions must always come in pairs, and these would generate 
the alternating group $A_5\cong PGL(2,4)$, the group of all even permutations of the
five labels $abcde$.
This is the group of `rotations' for $GQM(2,4)$.

The inequivalent dual-vectors of $V_4^*$ can be taken to be
\begin{equation}
\hspace{-0.2cm}
\begin{array}{rlrlrl}
\bra{\bar{a}} & =\; \bigl[\;\0\;\;\1     \;\bigr]\;, &
\bra{\bar{b}} & =\; \bigl[\;\1\;\;\0     \;\bigr]\;, &
\bra{\bar{c}} & =\; \bigl[\;\1\;\;\uom   \;\bigr]\;, \\
\bra{\bar{d}} & =\; \bigl[\;\1\;\;\uom^2 \;\bigr]\;, &
\bra{\bar{e}} & =\; \bigl[\;\1\;\;\1     \;\bigr]\;. & & \\
\end{array}
\end{equation}
%
The actions of these dual-vectors on the vectors are:
\smallskip
\begin{center}
\begin{tabular}{|r||l|l|l|l|l|}
\hline
& $\quad\ket{a}\quad$ & $\quad\ket{b}\quad$ & $\quad\ket{c}\quad$ & $\quad\ket{d}\quad$ & $\quad\ket{e}\quad$ \\
\hline
$\quad\bra{\bar{a}}\quad$ & $\;\quad\0$ & $\;\quad\1$     & $\;\quad\1$     & $\;\quad\1$     & $\;\quad\1$     \\
$\bra{\bar{b}}\quad$      & $\;\quad\1$ & $\;\quad\0$     & $\;\quad\uom$   & $\;\quad\uom^2$ & $\;\quad\1$     \\
$\bra{\bar{c}}\quad$      & $\;\quad\1$ & $\;\quad\uom$   & $\;\quad\0$     & $\;\quad\1$     & $\;\quad\uom^2$ \\
$\bra{\bar{d}}\quad$      & $\;\quad\1$ & $\;\quad\uom^2$ & $\;\quad\1$     & $\;\quad\0$     & $\;\quad\uom$   \\
$\bra{\bar{e}}\quad$      & $\;\quad\1$ & $\;\quad\1$     & $\;\quad\uom^2$ & $\;\quad\uom$   & $\;\quad\0$     \\
\hline
\end{tabular}
\end{center}
Thus,
\begin{eqnarray}
\braket{\bar{r}}{s} 
& =    & \0 \quad \mbox{if $r=s$,}     \cr
& \neq & \0 \quad \mbox{if $r\neq s$,}
\end{eqnarray}
and the relation $|\braket{\bar{r}}{s}|=1-\delta_{rs}$ is obtained
as before.  Maintaining this relation would require
relabeling the dual vectors in the same way as the vectors
under basis transformations.

\begin{figure}[b]
\caption{\label{Z4spins}To map `rotations' in $PGL(2,4)\cong A_5$ to rotations in $SO(3)$:
(a) label the faces of an icosahedron with five symbols as shown above left. 
Then, to every even permutation of the five labels
$abcde$ will correspond a rotation belonging to the icosahedral group.
(b) The `spin' observable $A_{ab}$ in $GQM(2,4)$ can be mapped onto a direction in 3D space as shown. 
(c) The urchin diagram showing all 20 `spin' directions of $GQM(2,4)$.\\}
\subfigure[]{\includegraphics[width=5.5cm]{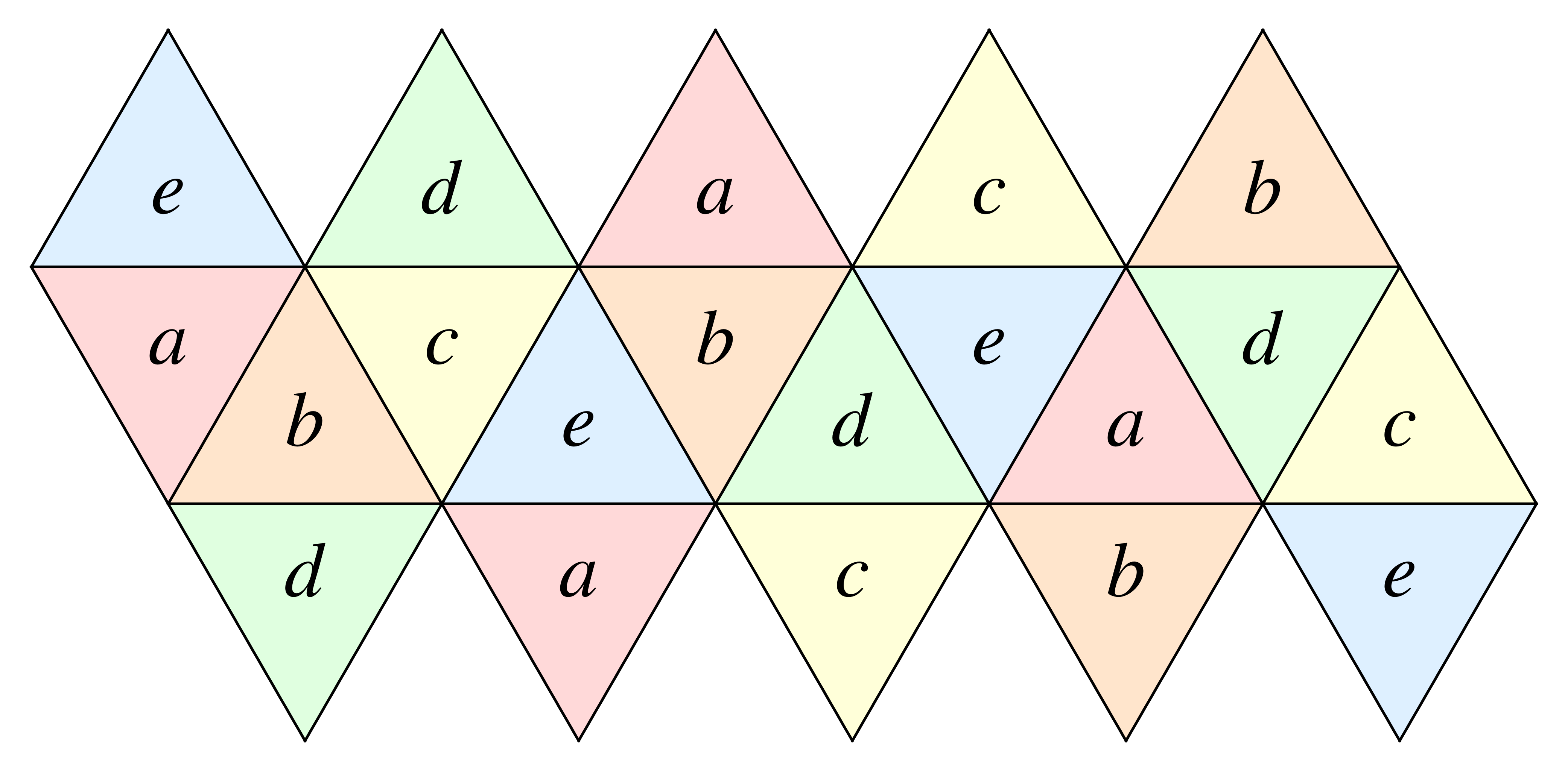}\hspace{0.0cm}}
\subfigure[]{\includegraphics[width=3.5cm]{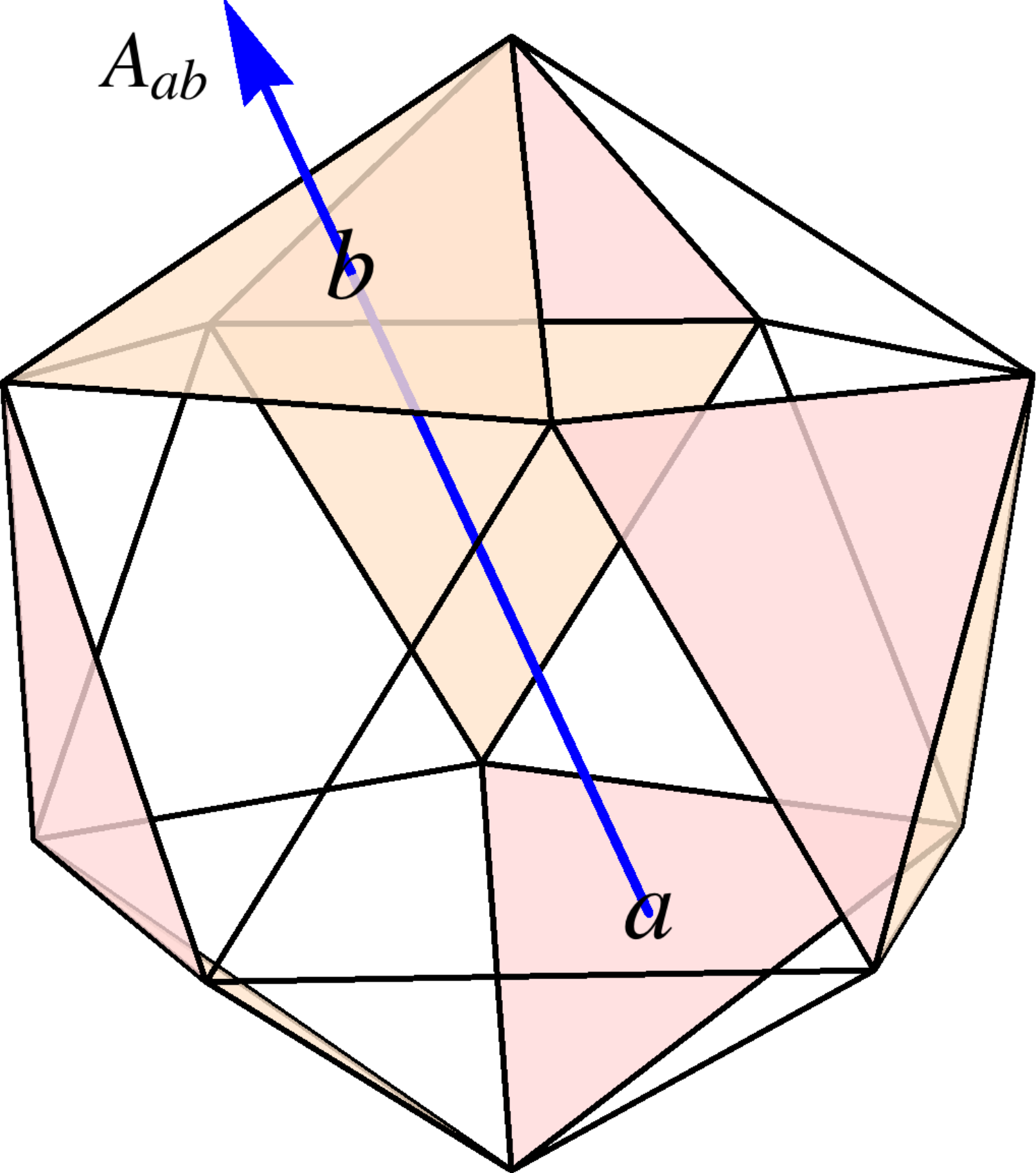}\hspace{0.2cm}}
\subfigure[]{\includegraphics[width=3.5cm]{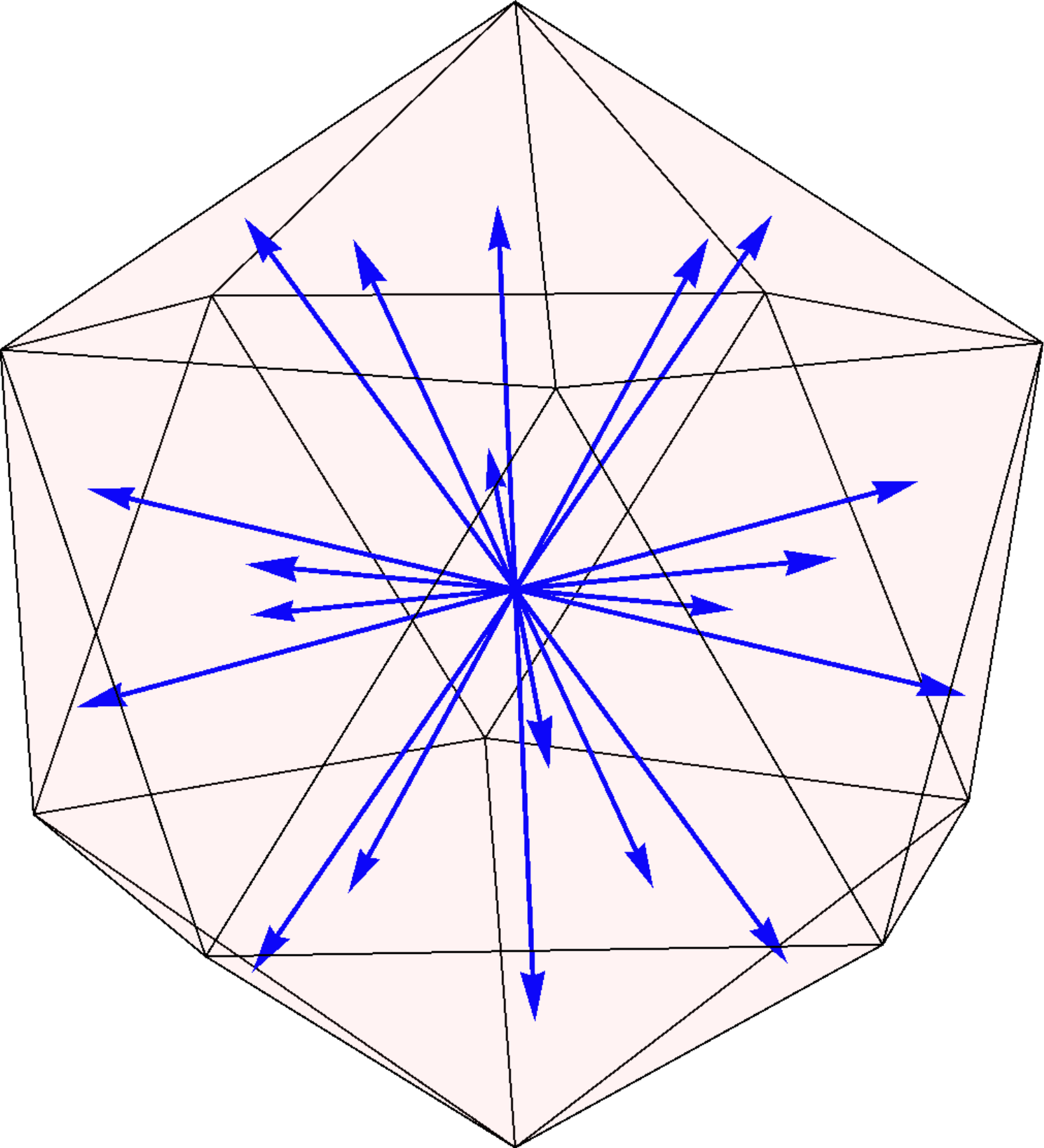}}
\end{figure}

From the set of five dual-vectors, we can define $5\times 4 = 20$ observables with outcomes
$\pm 1$, or 10 if we count the `spins' pointing in opposite directions as the same observable.
These `spins' can be associated with actual directions in 3D,
and their $PGL(2,4)\cong A_5$ transformations with rotations in $SO(3)$ as 
shown in \fref{Z4spins}:
First, label the faces of an icosahedron with five letters $abcde$, each 
letter appearing four time, as shown in \fref{Z4spins}(a).  The centers of the four faces
with the same letter are positioned at the vertices of a tetrahedron.
The icosahedral group $Y$ \cite{Hamermesh} which rotates the icosahedron onto itself consists of
60 elements.  Each of these elements will lead to an even permutation of the five letters on the
faces of the icosahedron.  Thus, there exists a one-to-one correspondence between
elements of the icosahedral group and the 60 permutations of $A_5$.  The `direction' of the `spin' 
observable $A_{ab}$ can be associated with the direction of the arrow shown in \fref{Z4spins}(b).
All 20 `spin' directions can be mapped this way, and \fref{Z4spins}(c) shows the
resulting urchin of spin-directions.
These 20 `spins' transform into each other under $A_5\cong Y$ rotations.

\subsection{$\mathbb{Z}_5$ case}

$GQM(2,5)$ is constructed on the field consisting of five elements,
$GF(5)=\mathbb{Z}_5=\mathbb{Z}/5\mathbb{Z}=\{\0,\1,\2,\3,\4\}$.
The addition and multiplication tables of this field is given by
\begin{center}
\begin{tabular}[t]{c|lllll}
$\;+\;$ & $\;\;\0\;$ & $\;\1\;$ & $\;\2\;$ & $\;\3\;$ & $\;\4\;$ \\
\hline
$\0$ & $\;\;\0$ & $\;\1$ & $\;\2$ & $\;\3$ & $\;\4$ \\
$\1$ & $\;\;\1$ & $\;\2$ & $\;\3$ & $\;\4$ & $\;\0$ \\
$\2$ & $\;\;\2$ & $\;\3$ & $\;\4$ & $\;\0$ & $\;\1$ \\
$\3$ & $\;\;\3$ & $\;\4$ & $\;\0$ & $\;\1$ & $\;\2$ \\
$\4$ & $\;\;\4$ & $\;\0$ & $\;\1$ & $\;\2$ & $\;\3$ 
\end{tabular}
\qquad
\begin{tabular}[t]{c|lllll}
$\;\times\;$ & $\;\;\0\;$ & $\;\1\;$ & $\;\2\;$ & $\;\3\;$ & $\;\4\;$ \\
\hline
$\0$ & $\;\;\0$ & $\;\0$ & $\;\0$ & $\;\0$ & $\;\0$ \\
$\1$ & $\;\;\0$ & $\;\1$ & $\;\2$ & $\;\3$ & $\;\4$ \\
$\2$ & $\;\;\0$ & $\;\2$ & $\;\4$ & $\;\1$ & $\;\3$ \\
$\3$ & $\;\;\0$ & $\;\3$ & $\;\1$ & $\;\4$ & $\;\2$ \\
$\4$ & $\;\;\0$ & $\;\4$ & $\;\3$ & $\;\2$ & $\;\1$ 
\end{tabular}
\end{center}
We will denote $\4=-\1$ and $\3=-\2$ in the following.
Since $\mathbb{Z}_5\backslash\{\0\}=\{\pm\1,\pm\2\}$, 
each physical state will be represented by four vectors in $V_5=\mathbb{Z}_5^2$
which differ by the multiplicative `phases' $-\1$ or $\pm\2$.

Thus, of the $5^2-1=24$ non-zero vectors in $V_5$, every four of them are equivalent, 
and the $24/4=6$ inequivalent ones can be taken to be:
\begin{equation}
\begin{array}{lll}
\ket{a} = \left[\begin{array}{r} \1   \\ \0 \end{array}\right],\quad &
\ket{b} = \left[\begin{array}{r} \0   \\ \1 \end{array}\right],\quad &
\ket{c} = \left[\begin{array}{r} \2   \\ \1 \end{array}\right],\quad \\
\ket{d} = \left[\begin{array}{r}-\1   \\ \1 \end{array}\right],\quad &
\ket{e} = \left[\begin{array}{r}-\2   \\ \1 \end{array}\right],\quad & 
\ket{f} = \left[\begin{array}{r} \1   \\ \1 \end{array}\right].    
\end{array}
\end{equation}
The group of basis transformations of this space is a subgroup of $S_6$, the group
of permutations of the six vector labels.  It consists of both odd and even permutations,
and the distribution of its elements among the 11 conjugate classes of $S_6$ are shown in
\tref{PGL25}.  There are 120 = 5! elements in total in 7 conjugate classes.
These numbers match those of $S_5$ exactly, and in fact, there is an isomorphism between the two,
\textit{i.e.} $PGL(2,5)\cong S_5$.  
Of the 120 elements of $PGL(2,5)$, a subgroup of 60 elements consisting of the
even permutations, and isomorphic to $A_5$, can be mapped onto $SO(3)$ rotations in the
icosahedral group $Y$ as we will see below.

\begin{table}[pt]
\caption{\label{PGL25}$PGL(2,5)$ is a subgroup of $S_6$, which is isomorphic to $S_5$. 
This table shows how many elements in each conjugate class of $S_6$ are in $PGL(2,5)$,
and the conjugate class in $S_5$ that they correspond to.
The signs adjacent to the Young tableaux indicate the signature of the
permutations in each class. Only the even permutations in $PGL(2,5)$, which form
an invariant subgroup of order 60 isomorphic to $A_5$, can be mapped to elements in $SO(3)$.\\}
\begin{indented}\item[]
\begin{tabular}{ccrrclc}
\br
\multicolumn{2}{c}{\ Conjugate\ \ }        &                       &                 & \multicolumn{2}{c}{\ Conjugate\ \ }        &       \\
\multicolumn{2}{c}{\ \ Classes of $S_6$\ \ \ } &  \hspace{0.8cm}$S_6$  & \hspace{0.5cm}$PGL(2,5)$  & \multicolumn{2}{c}{\ \ Classes of $S_5$\ \ \ } & $A_5$ \\ 
\br
& & & & & & \\
{\tiny$\yng(6)$}           & $-$ & 120 &  20\ \ \ \ \ & {\tiny$\yng(3,2)$}       & $-$ & \\
& & & & & & \\
{\tiny$\yng(5,1)$}         & $+$ & 144 &  24\ \ \ \ \ & {\tiny$\yng(5)$}         & $+$ & $\surd$ \\
& & & & & & \\
{\tiny$\yng(4,1,1)$}       & $-$ &  90 &  30\ \ \ \ \ & {\tiny$\yng(4,1)$}       & $-$ & \\
& & & & & & \\
{\tiny$\yng(4,2)$}         & $+$ &  90 &   0\ \ \ \ \ & & & \\
& & & & & & \\
{\tiny$\yng(3,3)$}         & $+$ &  40 &  20\ \ \ \ \ & {\tiny$\yng(3,1,1)$}     & $+$ & $\surd$ \\
& & & & & & \\
{\tiny$\yng(3,2,1)$}       & $-$ & 120 &   0\ \ \ \ \ & & & \\
& & & & & & \\
{\tiny$\yng(2,2,2)$}       & $-$ &  15 &  10\ \ \ \ \ & {\tiny$\yng(2,1,1,1)$}   & $-$ & \\
& & & & & & \\
{\tiny$\yng(2,2,1,1)$}     & $+$ &  45 &  15\ \ \ \ \ & {\tiny$\yng(2,2,1)$}     & $+$ & $\surd$ \\
& & & & & & \\
{\tiny$\yng(3,1,1,1)$}     & $+$ &  40 &   0\ \ \ \ \ & & & \\
& & & & & & \\
{\tiny$\yng(2,1,1,1,1)$}   & $-$ &  15 &   0\ \ \ \ \ & & & \\
& & & & & & \\
{\tiny$\yng(1,1,1,1,1,1)$} & $+$ &   1 &   1\ \ \ \ \ & {\tiny$\yng(1,1,1,1,1)$} & $+$ & $\surd$ \\
& & & & & & \\
\br
total                      &     & 720 & 120\ \ \ \ \ & & & \\
\br
\end{tabular}
\end{indented}
\end{table}

The inequivalent dual-vectors of $V_5^*$ can be taken to be
\begin{eqnarray}
\bra{\bar{a}} & = & \bigl[\;\0\;-\!\1           \;\bigr]\;,\quad
\bra{\bar{b}} \;=\; \bigl[\;\1\;\;\phantom{-}\0 \;\bigr]\;,\quad
\bra{\bar{c}} \;=\; \bigl[\;\1\;-\!\2           \;\bigr]\;,\cr
\bra{\bar{d}} & = & \bigl[\;\1\;\;\phantom{-}\1 \;\bigr]\;,\quad
\bra{\bar{e}} \;=\; \bigl[\;\1\;\;\phantom{-}\2 \;\bigr]\;,\quad
\bra{\bar{f}} \;=\; \bigl[\;\1\;-\!\1           \;\bigr]\;.
\end{eqnarray}
%
The actions of these dual-vectors on the vectors are:
\smallskip
\begin{center}
\begin{tabular}{|r||c|c|c|c|c|c|}
\hline
& $\quad\ket{a}\quad$ & $\quad\ket{b}\quad$ & $\quad\ket{c}\quad$ & $\quad\ket{d}\quad$ & $\quad\ket{e}\quad$ & $\quad\ket{f}\quad$ \\
\hline
$\quad\bra{\bar{a}}\quad$ & $\0$ & $-\1$           & $\phantom{-}\1$ & $-\1$           & $-\1$           & $-\1$ \\
$\bra{\bar{b}}\quad$      & $\1$ & $\phantom{-}\0$ & $\phantom{-}\2$ & $-\1$           & $-\2$           & $\phantom{-}\1$ \\
$\bra{\bar{c}}\quad$      & $\1$ & $-\2$           & $\phantom{-}\0$ & $\phantom{-}\2$ & $\phantom{-}\1$ & $-\1$ \\
$\bra{\bar{d}}\quad$      & $\1$ & $\phantom{-}\1$ & $-\2$           & $\phantom{-}\0$ & $-\1$           & $\phantom{-}\2$ \\
$\bra{\bar{e}}\quad$      & $\1$ & $\phantom{-}\2$ & $-\1$           & $\phantom{-}\1$ & $\phantom{-}\0$ & $-\2$ \\
$\bra{\bar{f}}\quad$      & $\1$ & $-\1$           & $\phantom{-}\1$ & $-\2$           & $\phantom{-}\2$ & $\phantom{-}\0$ \\
\hline
\end{tabular}
\end{center}
Thus,
\begin{eqnarray}
\braket{\bar{r}}{s} 
& =    & \0 \quad \mbox{if $r=s$,}     \cr
& \neq & \0 \quad \mbox{if $r\neq s$,}
\end{eqnarray}
and the relation $|\braket{\bar{r}}{s}|=1-\delta_{rs}$ is obtained
as before.  Maintaining this relation would require
relabeling the dual vectors in the same way as the vectors
under basis transformations.

\begin{figure}[b]
\caption{\label{Z5spins}Not all `rotations' in $PGL(2,5)\cong S_5$ can be mapped to rotations in $SO(3)$. 
The even permutations, isomorphic to $A_5$, can be mapped as follows:
(a) label the faces of a dodecahedron with six symbols as shown above left. 
Then, to every even permutation of the six labels
$abcdef$ will correspond a rotation belonging to the icosahedral group.
(b) The `spin' observable $A_{ab}$ in $GQM(2,5)$ can be mapped onto the direction in 3D space as shown.
(c) The urchin diagram showing all 30 `spin' directions of $GQM(2,5)$.\\}
\subfigure[]{\includegraphics[width=5cm]{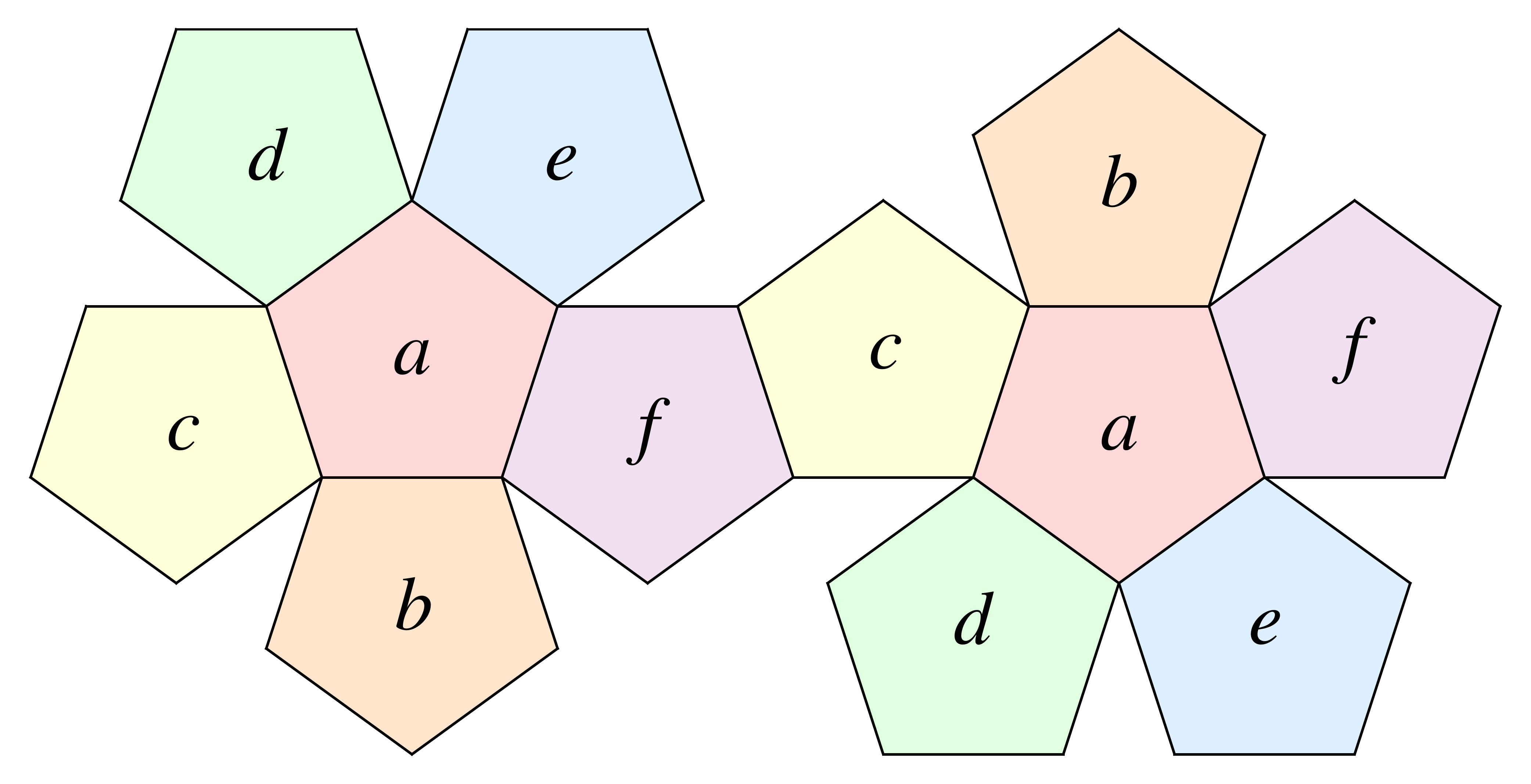}\hspace{0.1cm}}
\subfigure[]{\includegraphics[width=3.7cm]{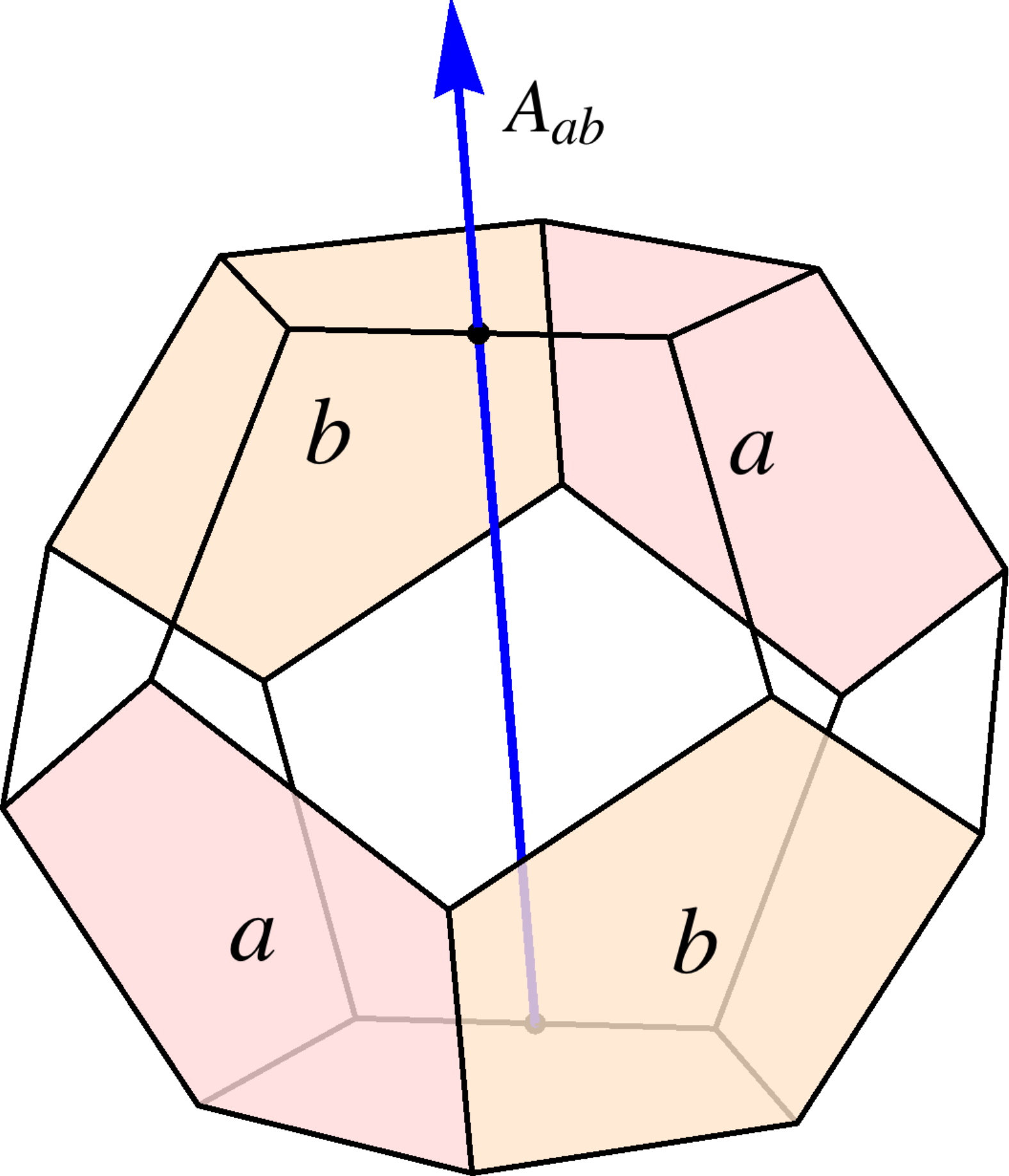}\hspace{0.1cm}}
\subfigure[]{\includegraphics[width=3.7cm]{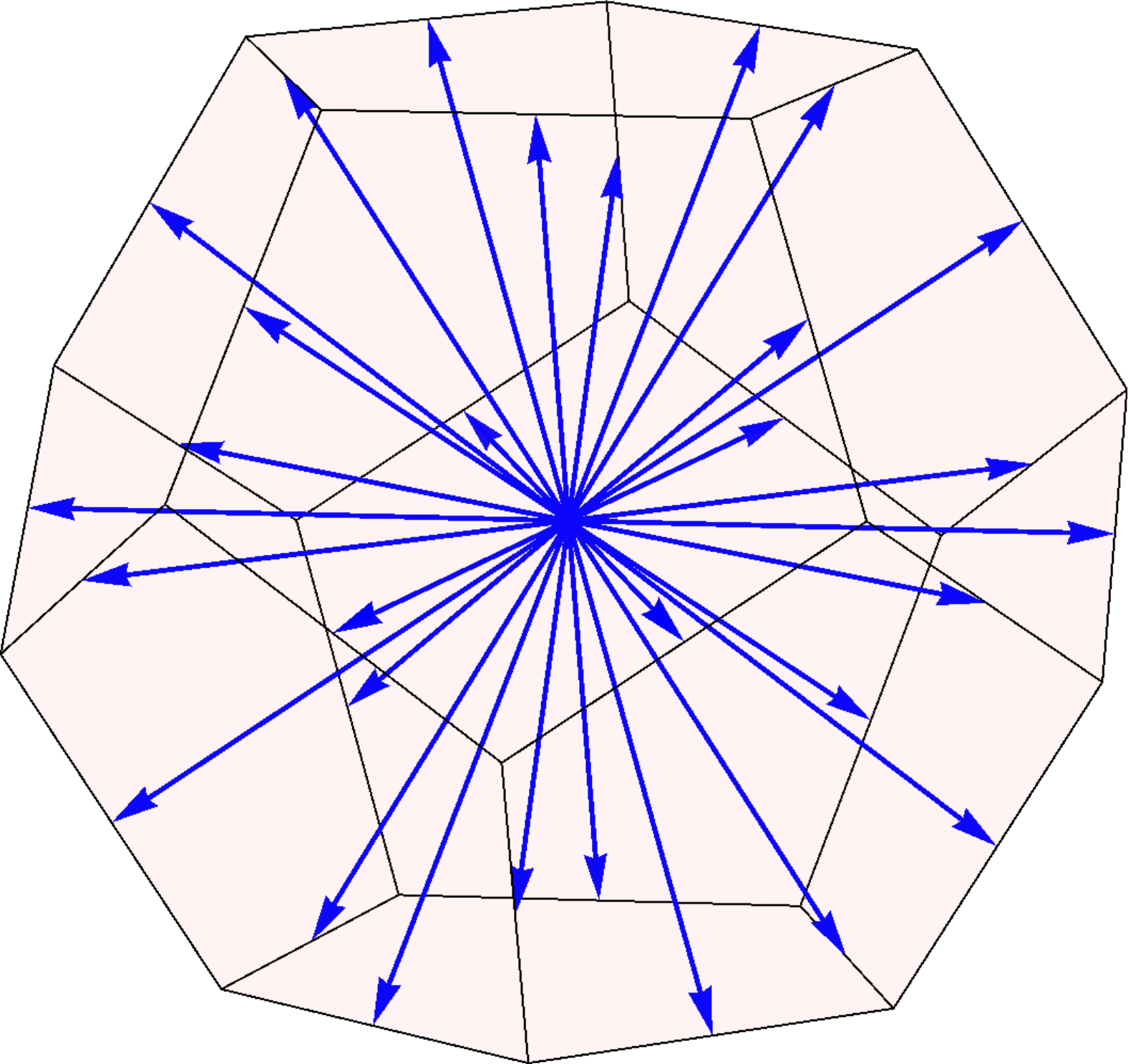}}
\end{figure}

From the set of six dual-vectors, we can define $6\times 5 = 30$ observables with outcomes
$\pm 1$, or 15 if we count the `spins' pointing in opposite directions as the same observable.
These `spins' can be associated with actual directions in 3D,
and their transformations under the subgroup of $PGL(2,5)$ mentioned above 
with rotations in $SO(3)$ as shown in \fref{Z5spins}:
First, label the 12 faces of a dodecahedron with 6 letters $abcdef$, with each letter 
appearing twice on faces that oppose each other, a shown in \fref{Z5spins}(a).
Note that the dodecahedron is dual to the icosahedron, under the interchange of vertices
and faces, so its symmetry group under rotations is the icosahedral group $Y$, which was shown
to be isomorphic to $A_5$ in the $q=4$ case we discussed above.
Each of the rotations of the icosahedral group will lead to an even permutation of the six letters on the
faces of the dodecahedron.  These will generate the subgroup of $PGL(2,5)$ consisting of even permutations only.
The `direction' of the `spin' 
observable $A_{ab}$ can be associated with the direction of the arrow shown in \fref{Z5spins}(b).
All 30 `spin' directions can be mapped this way, and \fref{Z5spins}(c) shows the
resulting urchin of spin-directions.
These 30 `spins' transform into each other under $A_5\cong Y$ rotations.

Unfortunately, there are 60 more elements of $PGL(2,5)$ unaccounted for, and these
do not seem to be representable as rotations or reflections of the dodecahedron.
Thus, in a sense, the `rotations' in the state space of $GQM(2,5)$ 
are much richer than a finite group of $SO(3)$ rotations.

\subsection{$\mathbb{Z}_7$ and Beyond}

As we have seen, for the $q=2$, $3$, $4$, and $5$ cases, the group $PGL(2,q)$ itself,
or its invariant subgroup, is isomorphic to some polyhedral group, allowing for the
identification of those $PGL(2,q)$ group elements with $SO(3)$ rotations.
Therefore, our `spins' can be considered objects that transform like canonical spin under
`rotations' for these cases.

Whether a similar pattern emerges for $GQM(2,7)$ and beyond remains to be explored.
In general, the $PGL(2,q)$ group is a subgroup of $S_{q+1}$ of order $q(q^2-1)$.
Constructing a correspondence via the method we employed in this paper would require
the labeling of the faces of some polyhedron with $q+1$ symbols.
Given that only 5 Platonic solids and 13 Archimedean solids are at our disposal,
it is not at all clear that such a correspondence exists for generic $q$.

\section{Summary}

In this paper, we have elaborated on Galois field quantum mechanics (GQM) introduced in our
previous letter \cite{Chang:2012}. In particular we have examined in detail the cases of
$GQM(2,2)$, $GQM(2,3)$, $GQM(2,4)$ and $GQM(2,5)$.
The fascinating geometric structure that underlies GQM as encapsulated by the
finite projective geometry \cite{Hirschfeld, Arnold, Ball-Weiner} is interesting in itself, 
and it also represents the simplest theoretical playground for understanding some outstanding issues 
in the foundations of quantum theory, quantum information and quantum computation \cite{James:2011} (see also \cite{Finkelstein:1996fn}).

It is not clear to us if GQM has an immediate realization in any physical system.    
Most dynamical systems exist in continuous spaces, with continuous group operations.    
However, there could exist parameter spaces of complex systems, 
such as those in quantum computation and information theory, 
which can be considered as finite fields, and in those cases, the results obtained here could have relevance.    
The pictures we present here could provide a context to better understand what is happening in these systems.

As already emphasized in \cite{Chang:2012}, GQM should also prove useful
in understanding the still mysterious super-quantum limit \cite{Chang:2011yt,super}
and its possible relation to quantum gravity \cite{Chang:2011yt}.  
That the Galois fields should be of relevance in going beyond quantum field theory
has been conjectured a long time ago \cite{nambu}, and our effort should be
understood as a natural realization of that prescient old intuition.

Features of GQM should shed light on questions raised in 
the geometric formulation of canonical quantum theory \cite{Ashtekar:1997ud} 
and in the natural generalization of the geometric quantum theory \cite{Jejjala:2007rn}
argued to be relevant to quantum gravity \cite{Minic:2003en}. 
These questions lie outside of the scope of this  paper, and will be taken up elsewhere \cite{next}.

\ack
We would like to thank Sir Anthony Leggett and Prof. Chia Tze for helpful discussions.  
ZL, DM, and TT are supported in part by
the U.S. Department of Energy, grant DE-FG05-92ER40677, task A.

\section*{References}

\end{document}